\documentclass[useAMS,usenatbib]{mn2e}
\usepackage{natbib}
\usepackage{amsmath,amsfonts}
\usepackage{epsfig}
\usepackage{mathptmx}

\def\reff@jnl#1{{\rm#1\/}}

\def\aj{\reff@jnl{AJ}}                  
\def\araa{\reff@jnl{ARA\&A}}            
\def\apj{\reff@jnl{ApJ}}                        
\def\apjl{\reff@jnl{ApJ}}               
\def\apjs{\reff@jnl{ApJS}}              
\def\ao{\reff@jnl{Appl.Optics}}         
\def\apss{\reff@jnl{Ap\&SS}}            
\def\aap{\reff@jnl{A\&A}}                       
\def\apjl{\reff@jnl{ApJ}}               
\def\aapr{\reff@jnl{A\&A~Rev.}}         
\def\aaps{\reff@jnl{A\&AS}}             
\def\azh{\reff@jnl{AZh}}                        
\def\baas{\reff@jnl{BAAS}}              
\def\jrasc{\reff@jnl{JRASC}}            
\def\memras{\reff@jnl{MmRAS}}           
\def\mnras{\reff@jnl{MNRAS}}            
\def\pra{\reff@jnl{Phys. Rev. A}}         
\def\prb{\reff@jnl{Phys. Rev. B}}         
\def\prc{\reff@jnl{Phys. Rev. C}}         
\def\prd{\reff@jnl{Phys. Rev. D}}         
\def\prl{\reff@jnl{Phys. Rev. Lett}}      
\def\pasp{\reff@jnl{PASP}}              
\def\pasj{\reff@jnl{PASJ}}              
\def\qjras{\reff@jnl{QJRAS}}            
\def\skytel{\reff@jnl{S\&T}}            
\def\solphys{\reff@jnl{Solar~Phys.}}    
\def\sovast{\reff@jnl{Soviet~Ast.}}     
\def\ssr{\reff@jnl{Space~Sci.Rev.}}     
\def\zap{\reff@jnl{ZAp}}                        
\def\nat{\reff@jnl{Nature}}             

\def\p#1by#2{{\partial{#1} \over \partial{#2}}}
\def\pp#1by#2#3{{\partial^2{#1} \over \partial{#2}\partial{#3}}}
\def\d#1by#2{{{\rm d}{#1} \over {\rm d}{#2}}}
\def\dd#1by#2#3{{{\rm d}^2{#1} \over {\rm d}{#2}{\rm d}{#3}}}

\title[AMI-LA observations of Perseus]{AMI-LA radio continuum observations of \emph{Spitzer} c2d small clouds and cores: Perseus region\thanks{We request that any reference to this paper cites ``AMI Consortium: Scaife et~al. 2011''.}}

\label{firstpage}

\author[Scaife et~al.]{
 AMI Consortium: Anna M. M. Scaife$^1$\thanks{email: ascaife@cp.dias.ie},
 Jennifer Hatchell$^2$, 
 Matthew Davies$^3$,
\newauthor
 Thomas M. O. Franzen$^3$,
 Keith J. B. Grainge$^{3,4}$,
 Michael P. Hobson$^3$,
 Natasha Hurley-Walker$^3$,
\newauthor
 Anthony N. Lasenby$^{3,4}$,
 Malak Olamaie$^3$,
 Yvette C. Perrott$^3$,
 Guy G. Pooley$^3$,
\newauthor
 Carmen Rodr{\'i}guez-Gonz{\'a}lvez$^3$,
 Richard D. E. Saunders$^{3,4}$,
 Michel P. Schammel$^3$,
\newauthor
 Paul F. Scott$^3$,
 Timothy Shimwell$^3$,
 David Titterington$^3$,
 Elizabeth Waldram$^3$.
\vspace{0.03in}\\
$^1$ Dublin Institute for Advanced Studies, 31 Fitzwilliam Place,
     Dublin 2, Ireland\\
$^2$ School of Physics, University of Exeter, Stocker Road, Exeter EX4 4QL\\
$^3$ Astrophysics Group, Cavendish Laboratory, J J Thomson Avenue,
     Cambridge CB3 0HE\\
$^4$ Kavli Institute for Cosmology, Cambridge, Madingley Road,
     Cambridge, CB3 0HA\\
}

\date{Accepted ---; received ---; in original form \today}

\pagerange{\pageref{firstpage}--\pageref{lastpage}}

\pubyear{2010}

\begin{document}
\maketitle

\begin{abstract}
We present deep radio continuum observations of the cores identified as deeply embedded young stellar objects in the Perseus molecular cloud by the \emph{Spitzer} c2d programme at a wavelength of 1.8\,cm with the Arcminute Microkelvin Imager Large Array (AMI-LA). We detect 72\% of Class 0 objects from this sample and 31\% of Class I objects. No starless cores are detected. We use the flux densities measured from these data to improve constraints on the correlations between radio luminosity and bolometric luminosity, infra-red luminosity and, where measured, outflow force. We discuss the differing behaviour of these objects as a function of protostellar class and investigate the differences in radio emission as a function of core mass. Two of four possible very low luminosity objects (VeLLOs) are detected at 1.8\,cm.
\end{abstract}


\section{Introduction}

Since the advent of the \emph{Spitzer} space telescope, direct infrared detection of the hot dust heated by an embedded protostar has become the most popular method for differentiating between starless and protostellar cores. However, \emph{Spitzer} measurements of the luminosity distribution of protostars (Dunham et~al.  2008) have aggravated the `luminosity problem' first articulated by Kenyon et~al. (1990). This problem arises from the increasing number of protostellar objects being discovered with internal luminosities $L_{\rm{int}} < 1$\,L$_{\odot}$, and furthermore $L_{\rm{int}} < 0.1$\,L$_{\odot}$ in the case of very low luminosity objects (VeLLOs). Such low luminosities violate the lower luminosity limits set by steady accretion models for young stellar objects (Evans et~al. 2009; Shu 1977) and imply that alternative processes, such as non-steady accretion (Kenyon \& Hartmann 1995; Young \& Evans 2005; Enoch et~al. 2007) are required to explain this discrepancy. This luminosity problem is most difficult to rectify in very low luminosity objects (VeLLOs; Young et~al. 2004; Dunham et~al. 2008) with extreme luminosities $L_{\rm{int}}\leq 0.1~L_{\odot}$. The nature of these objects is unclear, whether they are young Class 0 protostars which are just powering up, or are more evolved but in a low accretion state (Dunham et~al. 2008; Evans et~al. 2009). 

Although the method is extremely reliable, \emph{Spitzer} identification is not completely certain (Hatchell \& Dunham 2009) as the response to extra-galactic radio galaxies is known to mimic that of embedded protostellar cores and can cause false detections. In addition, a protostellar spectrum alone will not demonstrate that a core is truly embedded and, like many surveys, \emph{Spitzer} is also flux-limited, with a luminosity completeness limit of $0.004(d/{140~{\rm pc}})^2$  ${L}_{\odot}$ (Dunham et al. 2008; Harvey et al. 2007). Correctly determining the relative numbers of starless and protostellar cores in star-forming regions is essential for inferring timescales for the different stages of protostellar evolution; for low-luminosity objects it is also necessary in order to determine the extent of the luminosity problem. Identifying embedded protostars by detecting their molecular outflows, either via high-excitation jet interactions (Herbig-Haro objects and shock-excited H$_2$; eg. Davis et al. 2008; Walawender et al. 2005) or through low-excitation molecular lines such as $^{12}$CO, can immediately identify a source as an embedded protostar, avoiding the contaminants of infrared colour selection. However, in crowded star-formation regions, confusion from neighbouring outflows can frequently be an issue, as we discuss later. Outflows are particularly interesting for VeLLO scenarios by identifying if the protostar has shown more active mass ejection in the past, as momentum deposited at earlier times remains visible in the molecular outflow.

A further method for detecting protostars is through their radio emission (see e.g. Andr{\'e}, Ward-Thompson \& Barsony 1993). This radio emission provides not only a detection method, but also a potential mechanism for distinguishing protostellar class via its correlations with other physical characteristics. Radio follow-up of the Dunham et~al. (2008) \emph{Spitzer} catalogue of low luminosity embedded objects at 16\,GHz (AMI Consortium: Scaife et~al. 2010) has provided clear evidence for distinct trends in the behaviour of the radio luminosity of these sources when compared with their bolometric luminosity, IR luminosity and outflow force where molecular outflows have been measured. Similar trends are well-documented at lower radio frequencies for higher luminosity sources ($<10^3$\,L$_{\odot}$), see e.g. Anglada (1995) or Shirley et~al. (2007). In addition, the relationship between radio luminosity and the IR luminosity of these objects has shown that it may be possible to use the radio emission as a proxy for the internal luminosity of low luminosity objects, a quantity which can otherwise only be derived through complex modelling of the IR spectra.

Here we present new data following on from the initial work of AMI Consortium: Scaife et~al. (2010), hereafter Paper I. The candidates identified by \emph{Spitzer} data in the catalogue of Dunham et~al. (2008) were ranked as belonging to one of six ``groups'', with Group 1 being those most likely to be true embedded objects, and Group 6 those least likely. These new data cover all objects classified as Group 1--3 in Dunham et~al. (2008) at $\delta\geq 15^{\circ}$ not covered by Paper I. These sources lie predominantly in the area of the Perseus molecular cloud and significantly increase the available high frequency radio data for this region. The Perseus molecular cloud is an exceptionally well studied star formation region with extensive complementary multi-frequency data available (e.g. Hatchell et~al. 2005; 2007a;b; Hatchell \& Dunham 2009). The sources investigated here all form part of the low- and very low luminosity object population which has aggravated the `luminosity problem'. As such the detailed correlations and derivations of their physical properties are of particular interest.

The organization of this paper is as follows. In \S~\ref{sec:sample} we describe the sample of targets to be observed, and in \S~\ref{sec:obs} we describe the AMI Large Array (AMI-LA) telescope, the observations and the data reduction process. In \S~\ref{sec:results} we comment upon the results of the observations and compare them to predictions. The nature of the radio emission and the correlations derived in \S~\ref{sec:pstars} are discussed in \S~\ref{sec:disc}. 

\section{The Sample}
\label{sec:sample}
Previous centimetre-wave radio continuum follow-up (Paper I) of the \emph{Spitzer} c2d catalogue of deeply embedded protostars (Dunham et~al. 2008) covered all those objects which lay within clouds targeted by the AMI-SZA spinning dust sample (Scaife et~al. in prep). Candidates for embedded objects from this catalogue were ranked by Dunham et al. (2008) as belonging to one of six ``groups'', with Group~1 being those most likely to be true embedded objects, and Group~6 those least likely.
The sample observed in this paper targets all the remaining candidates not observed in Paper I and identified as belonging to Groups 1-3 in Dunham et~al. (2008), although we note that no Group~2 candidates are present, above $\delta=15^{\circ}$. This declination limit is a consequence of the observing range of the AMI telescope; below this declination significant interference from geostationary satellites is experienced. The objects in this sample are predominantly found in the Perseus molecular cloud (22 cores) with a small number being located elsewhere (6 cores). We identify the candidates from this catalogue by their catalogue number, i.e. [DCE08]-nnn, in column [1] of Table~\ref{tab:sample}. The coordinates of the individual candidates are listed in columns [2] and [3] of Table~\ref{tab:sample}, along with their Group and infrared luminosity from Dunham et~al. (2008) in columns [4] and [5]. They have been cross-referenced with the sub-mm star formation survey of the Perseus molecular cloud (Hatchell et~al. 2005; Hatchell et~al. 2007a, hereafter H07), the designations from which are listed in column [7] and the Class assigned to each object from those works is listed in column [8].
\begin{table*}
\caption{The AMI-LA sample of embedded protostellar sources selected from the catalogue of Dunham et~al. (2008). Columns are [1] source number from the Dunham et~al. (2008) catalogue; [2] Right Ascension of source in J2000 coordinates; [3] Declination of source in J2000 coordinates; [4] candidate Group from Dunham et~al. (2008); [4] IR luminosity from Dunham et~al. (2008); [5] distance in kiloparsecs; [6] cross-identification with the catalogue of Hatchell et~al. (2007); and [7] protostellar Class, where `S' indicates a starless core. \label{tab:sample}}
\begin{tabular}{cccccccc}
\hline\hline
[DCE08] & RA & Dec. & Group & $L_{\rm{IR}}$ & $D$ & [H07] & Class \\
 & (J2000) & (J2000) & & (L$_{\odot}$) & (kpc) & & \\
\hline
055& 03 25 36.22 &+30 45 15.8  &1  &   0.061	&0.25	&28	&0\\
056& 03 25 39.12 &+30 43 58.1  &1  &   0.432	&0.25   &29	&0\\
063& 03 27 38.26 &+30 13 58.8  &1  &   0.347	&0.25   &39	&I\\
064& 03 28 32.57 &+31 11 05.3  &1  &   0.036	&0.25   &74	&I\\
065& 03 28 39.10 &+31 06 01.8  &1  &   0.015    &0.25   &71	&0\\
068& 03 28 45.29 &+31 05 42.0  &1  &   0.185    &0.25   &-	&-\\
071& 03 29 00.55 &+31 12 00.7  &1  &   0.105    &0.25   &65	&0\\
073& 03 29 12.07 &+31 13 01.6  &1  &   0.018 	&0.25   &42	&0\\
081& 03 30 32.69 &+30 26 26.5  &1  &   0.033	&0.25   &Bolo62$^a$	&0\\
084& 03 31 20.98 &+30 45 30.2  &1  &   0.299    &0.25   &77	&0\\
088& 03 32 17.95 &+30 49 47.6  &1  &   0.135    &0.25   &76	&0\\
090& 03 32 29.18 &+31 02 40.9  &1  &   0.180    &0.25   &-	&-\\
092& 03 33 14.38 &+31 07 10.9  &1  &   0.098    &0.25   &-	&-\\
093& 03 33 16.44 &+31 06 52.6  &1  &   0.144    &0.25   &4	&0\\
105& 03 43 56.52 &+32 00 52.9  &1  &   0.310    &0.25   &12	&0\\
106& 03 43 56.83 &+32 03 04.7  &1  &   0.253    &0.25   &13	&0\\
\hline   
060& 03 26 37.46 &+30 15 28.1  &3  &   0.462    &0.25   &80	&I\\
080& 03 29 51.82 &+31 39 06.1  &3  &   0.166    &0.25   &-	&-\\
104& 03 43 51.02 &+32 03 07.9  &3  &   0.172    &0.25   &15	&0\\
107& 03 44 02.40 &+32 02 04.9  &3  &   0.101    &0.25   &16/18?	&S/S\\
108& 03 44 02.64 &+32 01 59.5  &3  &   0.021    &0.25   &16/18?	&S/S\\
109& 03 44 21.36 &+31 59 32.6  &3  &   0.200    &0.25   &-	&-\\
\hline\hline
005& 04 41 12.65 &+25 46 35.4  &1  &   0.383	&0.14	&-      &-\\
044& 22 30 31.94 &+75 14 08.9  &1  &   0.156   	&0.30	&-      &-\\
045& 22 31 05.59 &+75 13 37.2  &1  &   0.077 	&0.30	&-      &-\\
048& 22 38 46.15 &+75 11 32.3  &1  &   0.274 	&0.30	&-      &-\\
049& 22 38 46.44 &+75 11 28.0  &1  &   0.081 	&0.30	&-      &-\\
\hline
043& 22 29 59.52 &+75 14 03.1  &3  &   0.272 	&0.30	&-      &-\\
\hline
\end{tabular}
\begin{minipage}{\textwidth}{
$^a$ original designation from Enoch et~al. 2007.
}
\end{minipage}
\end{table*}

\section{Observations}
\label{sec:obs}

AMI comprises two synthesis arrays, one of ten 3.7\,m
antennas (SA) and one of eight 13\,m antennas (LA),
both sited at the Mullard Radio Astronomy Observatory at Lord's Bridge, Cambridge (AMI Consortium: Zwart
et~al. 2008). The telescope observes in 
the band 13.5--17.9\,GHz with eight 0.75\,GHz bandwidth channels. In practice, the two
lowest frequency channels (1 \& 2) are not generally used due to a lower response in this frequency range and interference from geostationary
satellites. 

Observations of the 28 objects listed in Table~\ref{tab:sample} were made with the AMI-LA between July 2010 and August 2010. Each target was observed as a single pointing, with the exceptions of [DCE08]-048 which is located within the [DCE08]-049 pointing (separation$\approx4"$), and [DCE08]-107 which is within the [DCE08]-108 pointing (separation$\approx6"$).
\begin{table}
\caption{AMI-LA frequency channels and primary calibrator flux densities measured in Jy.\label{tab:cals}}
\begin{tabular}{lcccccc}
\hline \hline
Channel No. & 3 & 4 & 5 & 6 & 7 & 8\\
Freq. [GHz] & 13.88 & 14.63 & 15.38 & 16.13 & 16.88 & 17.63 \\
\hline
3C48  & 1.89 & 1.78 & 1.68 & 1.60 & 1.52 & 1.45 \\
3C286 & 3.74 & 3.60 & 3.47 & 3.35 & 3.24 & 3.14 \\
3C147 & 2.85 & 2.71 & 2.57 & 2.45 & 2.35 & 2.24 \\
\hline
\end{tabular}
\end{table}

AMI-LA data reduction is performed using the local software tool \textsc{reduce}. This applies
both automatic and manual flags for interference, 
shadowing and hardware errors, Fourier transforms the lag-delay correlator data to synthesize frequency
channels and performs phase and amplitude
calibrations before output to disc in \emph{uv} FITS format suitable for imaging in
\textsc{aips}\footnote{\tt http://www.aips.nrao.edu/}. Flux (primary) calibration is performed using short observations of 3C286 and 3C48. We assume I+Q flux densities for this source in the
AMI LA channels consistent with the updated VLA calibration scale (Rick Perley, private comm.), see Table~\ref{tab:cals}. Since the AMI-LA measures
I+Q, these flux densities 
include corrections for the polarization of the calibrator sources. A correction is
also made for the changing intervening air mass over the observation. From
other measurements, we find the flux calibration is accurate to better than
5 per cent (AMI Consortium: Scaife et~al. 2008; AMI Consortium:
Hurley--Walker et~al. 2009). Additional phase (secondary) calibration is done using interleaved observations of
calibrators 
selected from the Jodrell Bank VLA Survey (JVAS; Patnaik et~al. 1992). After calibration, the phase is generally stable to
$5^{\circ}$ for channels 4--7, and
$10^{\circ}$ for channels 3 and 8. The FWHM of the primary beam of the AMI LA is $\approx 6$\,arcmin at 16\,GHz. The data in this paper were taken with the AMI Large Array (AMI-LA) using channels $4-7$, which have superior phase stability. Therefore measured flux densities in what follows are made with a total bandwidth of 3\,GHz.

Reduced data were imaged using the AIPS data package. {\sc{clean}}
deconvolution was performed using the task 
{\sc{imagr}} which applies a differential primary beam correction to
the individual frequency channels to produce the combined frequency
image.  In what
follows we use the convention: $S_{\nu}\propto \nu^{\alpha}$, where $S_{\nu}$ is
flux density (rather than flux, $F_{\nu}=\nu S_{\nu}$), $\nu$ is frequency and $\alpha$ is the spectral index. All errors quoted are 1\,$\sigma$. 
\begin{table}
\caption{AMI LA observations. Sources found within the Perseus region are listed first, with those outside Perseus continued below the double line division. Columns are [1] source number from the Dunham et~al. (2008) catalogue; [2] AMI-LA flux calibrator; [3] AMI-LA phase calibrator; [4] rms noise measured from recovered map; [5] major axis of AMI-LA synthesized beam; [6] minor axis of AMI-LA synthesized beam.  \label{tab:obs}}
\begin{tabular}{cccccc}
\hline\hline
[DCE08] & $1^{\circ}$ & $2^{\circ}$ & $\sigma_{\rm{rms}}$ & $\Delta \theta_{\rm{max}}$ & $\Delta \theta_{\rm{min}}$ \\
&&&($\frac{\mu \rm{Jy}}{\rm{beam}}$) & (arcsec) & (arcsec) \\
\hline
055& 3C286& J0329+2756 & 19  & 41.4 & 28.2 \\
056& 3C286& J0329+2756 & 22  & 41.8 & 28.6 \\
063& 3C48 & J0329+2756 & 25  & 45.0 & 26.4 \\
064& 3C48 & J0329+2756 & 18  & 40.5 & 26.3 \\
065& 3C286& J0329+2756 & 20  & 40.2 & 29.8 \\
068& 3C48 & J0329+2756 & 18  & 42.7 & 27.5 \\
071& 3C48 & J0329+2756 & 19  & 36.6 & 23.9 \\
073& 3C48 & J0329+2756 & 26  & 37.3 & 24.7 \\
081& 3C48 & J0329+2756 & 20  & 39.4 & 29.0 \\
084& 3C48 & J0329+2756 & 30  & 36.3 & 25.2 \\
088& 3C48 & J0329+2756 & 32  & 36.9 & 24.4 \\
090& 3C48 & J0329+2756 & 25 & 36.1 & 26.1 \\
092& 3C48 & J0329+2756 & 27 & 36.8 & 24.6 \\
093& 3C48 & J0329+2756 & 22 & 36.4 & 24.5 \\
105& 3C48 & J0329+2756 & 23 & 38.6 & 24.8 \\
106& 3C48 & J0329+2756 & 20 & 41.0 & 29.7 \\
\hline
060& 3C48 & J0329+2756 & 19 & 46.0 & 25.5 \\
080& 3C48 & J0329+2756 & 19 & 42.5 & 28.4 \\
104& 3C48 & J0341+3352 & 19 & 36.9 & 24.2 \\
107& 3C48 & J2302+6405 & 22 & 41.9 & 29.0 \\
108& 3C48 & J2302+6405 & 22 & 41.9 & 29.0 \\
109& 3C48 & J0341+3352 & 25  & 35.8 & 23.7 \\
\hline\hline
005& 3C286& J0435+2532 & 25  & 43.2 & 29.2 \\
044& 3C286& J2236+7322 & 24  & 40.5 & 25.3 \\
045& 3C286& J2236+7322 & 15  & 36.7 & 27.5 \\
048& 3C48 & J2236+7322 & 25  & 30.8 & 24.7 \\
049& 3C286& J2236+7322 & 25  & 30.8 & 24.7 \\
\hline
043& 3C286& J2236+7322 & 24  & 40.7 & 28.0 \\
\hline
\end{tabular}
\end{table}

\section{Results}
\label{sec:results}

The observations towards the 28 targets listed in Table~\ref{tab:sample} and described in \S~\ref{sec:sample} are summarized in Table~\ref{tab:obs}. The details of each individual observation including both the primary and secondary calibration sources are listed. Maps were made using naturally weighted visibilities to ensure optimal noise levels, except in those cases where sources were separated by less than one synthesised beam. In these instances uniform weighting was used in order to improve resolution and separate the sources. The resulting rms noise level in each map and the dimensions of the synthesized beam are listed in Table~\ref{tab:obs}, and where uniform weighting has been employed it is indicated in the table. The rms noise level varies between fields due to the different levels of data flagging required following periods of poor weather conditions or interference from non-astronomical sources, such as geostationary satellites. 

Source detection for these data was performed in the un-primary-beam-corrected maps, where we identified all objects within the FWHM of the AMI-LA primary beam with a peak flux density $>5\,\sigma_{\rm{rms}}$, as being true sources. A full list of the sources detected in these fields is given in Tables~\ref{tab:moresources1} \&~\ref{tab:moresources2}. The peak and integrated flux densities listed in both Table~\ref{tab:phys} and Tables~\ref{tab:moresources1} \&~\ref{tab:moresources2} were extracted from the primary-beam-corrected maps. 

Integrated flux densities were determined by fitting a multivariate Gaussian and a background level to each source using the standard {\sc aips} task {\sc jmfit}. In addition, those objects which were poorly fitted by a Gaussian source model have integrated flux densities found using the {\sc fitflux} program (Green 2007). This method calculates flux densities by removing a tilted plane fitted to the local background and integrating the remaining flux density. This is done by drawing a polygon around the source and fitting a tilted plane to the pixels around the edge of the polygon. Where an edge of the polygon crosses a region confused by another source, the background is subjective and this edge is omitted from the fitting. Since the extracted flux density is dependent to some degree on the background emission, we repeated this process using five irregular polygons, each varying slightly in shape. The final flux density is the average of that extracted from these five polygons. The error on these integrated flux densities is calculated as $\sigma_{\rm{S}}=\sqrt{(0.05S_{\rm{int}})^2+\sigma^2_{\rm{fit}}+\sigma^2_{\rm{rms}}}$, where $\sigma_{\rm{fit}}$ is the standard deviation of the flux densities found from the five polygonal apertures, $\sigma_{\rm{rms}}$ is the rms noise determined from the maps  and $0.05S_{\rm{int}}$ is a conservative 5 per cent absolute calibration error. In the case of sources fitted using {\sc jmfit} $\sigma_{\rm{fit}}$ is the fitting error returned by the task. Where the 5 per cent calibration uncertainty is not dominant, errors determined for this sample are in general heavily dependent on the rms noise, $\sigma_{\rm{rms}}$, rather than the fitting error, $\sigma_{\rm{fit}}$, which is found to be small.

The source list in Tables~\ref{tab:moresources1} \&~\ref{tab:moresources2} was cross referenced with the literature and where the position of a source coincides with a known object this has been indicated in Column [7]. For this identification we employed a search radius of $30''$, although we note that most associations are within $<2''$. The only source significantly separated from its association is [DCE08]-068 and this is commented upon in later discussions. Sources are identified as extragalactic radio sources where they have a counterpart in the NVSS (Condon et~al. 1998) catalogue, and are denoted ``radio". 

\begin{table*}
\caption{Physical parameters of protostellar sources within Perseus. Columns are [1] source number from the H07 catalogue; [2] source number from the DCE08 catalogue; [3] protostellar Class as identified in H07; [4] flux density predicted from vibrational dust emission at 16\,GHz; [5] flag on presence of radio counterpart; [6] 1.8\,cm radio flux density in excess of the thermal greybody emission measured by the AMI-LA; [7] envelope mass from H07 corrected to $D=250$\,kpc; [8] Bolometric temperature from H07; [9] Bolometric luminosity from H07 corrected to $D=250$\,kpc, unless otherwise indicated; [10] Outflow force from H07b corrected to $D=250$\,kpc; [11] flag on possible confusion of the outflow with other nearby objects. \label{tab:phys}}
\begin{tabular}{ccccccccccc}
\hline \hline
[H07] & [DCE08] & Class & $S_{\rm{pred,16}}$ &radio det.& $S_{16}$ & $M_{\rm{env}}$ & $T_{\rm{bol}}$ & $L_{\rm{bol}}$ & $F_{\rm{out}}$ & confused?\\
& & & ($\mu$Jy) & (y/n) & ($\mu$Jy) & (M$_{\odot}$) & (K) & (L$_{\odot}$) & ($10^{-5}\,$M$_{\odot}\,$yr$^{-1}$\,km\,s$^{-1}$) & \\
\hline
1 & & 0 & 88 &y& 336.2 & 10.8 & 53 & 2.3 & $1.3\pm0.8$ &\\
2 & & 0 &140&y& 499.9 & 15.9 & $<25$ & $<1.5$ & $9.6\pm1.0$ &\\
4 & 093 & 0 &51 &y& 191.6 & 9.2 & 32 & 1.1$^{\ast}$ & $1.4\pm0.3$ &\\
7 & & I &9  &y& 128.5 & 1.7 & 158 & 0.8 & $9.8\pm0.8$ &\\
10& & I &3  &y& 578.9 & 0.6 & 117 & 0.2 & $3.4\pm0.4$ &\\
12& 105 & 0 &71 &y& 911.1 & 14.0 & 31 & 1.8$^{\ast}$ & $4.8\pm0.8$ &\\
13& 106 & 0 &61 &y& 490.6 & 6.2 & 59 & 1.1$^{\ast}$ & $4.0\pm0.5$ &\\
15& 104 & 0 &12 &n& $<93$ & 2.7 & 67 & 0.32$^{\ast}$ & $0.5\pm0.2$ &\\
16/18 & 107 & S & 16$^c$ & n & $<110$ & - & 76$^{\ast}$ & 0.18$^{\ast}$ & - &  \\
16/18 & 108 & S & 16$^c$ & n & $<110$ & - & -  & -    & - &  \\
27& & 0 &128&y& 1856.8$^c$& 18.8 & 53 & 3.3 & $21.5\pm3.0$&\\
28& 055 & 0 &67 &y& 1858.7$^c$& 9.9 & 53 & 10.4 & $28.4\pm2.9$&\\
29& 056 & 0 &55 &y& 473.0 & 11.1 & 49 & 3.8 & $16.3\pm1.5$&\\
30& & 0 &60 &y& 610.2 & 10.0 & 50 & 2.5 & $7.9\pm0.8$&\\
31& & 0 &59 &n& $<140$ & 7.3 & 38 & 1.2 & $2.8\pm0.5$ & c \\
35& & I &17 &n& $<1135$& 3.2 & 105 & 4.3 & $7.3\pm0.6$&\\
36& & 0 &31 &y& $-^{d}$    & 3.2 & 59 & 1.9 & $5.1\pm0.5$&\\
37& & I &27 &n& $<98$ & 4.3 & 672 & 2.0 & $2.5\pm0.4$ & c \\
39& 063 & I &9(27$^a$)  &n& $<116(98)$ & 1.2 & 394 & 1.1$^{\ast}$ & $1.4\pm0.2$ & c \\
41& & 0 &231&y& 1587.0& 31.8 & 35 & 5.2 & $35.5\pm2.9$&\\
42& 073 & 0 &77 &y& 391.3 & 15.4 & 39 & 4.6 & $8.1\pm1.6$&\\
44& & 0 &87 &y& 233.4 & 16.6 & 58 & 13.8 & $56.7\pm4.4$&\\
48& & 0 &36 &n& $<236$ & 5.0 & 49 & 2.0 & $2.1\pm1.6$ & c \\
49& & I &28 &n& $<217$ & 4.5 & 140 & 4.1 & H&\\
50& & I &41 &n& $<245$ & 4.4 & 88 & 3.4 & $129.6\pm10.9$ & c\\
52& & 0 &33 &y& $-^{d}$ & 3.7 & 57 & 2.7 & $25.4\pm2.6$ & c \\
65& 071 & 0 &12(33$^a$) &y& 227.7(206.7)& 1.2 & 56 & 0.24$^{\ast}$ & - &\\
69& & I &3  &n& $<116$ & 0.4 & 436 & 0.02 & H?&\\
71& 065 & 0 &12 &y& 275.6 & 1.3 & 37 & 0.22$^{\ast}$& H?&\\
74& 064 & I &3  &n& $<87$ & 0.4 & 144 & 0.20$^{\ast}$& H&\\
75& & 0 &4  &n& $<96$ & 0.7 & 67 & 0.06 & H&\\
76& 088 & 0 &104&y& 320.2 & 9.6 & 109 & 1.2$^{\ast}$ & $0.9\pm0.5$&\\
77& 084 & 0 &19(37$^a$) &y& 305.1(287.1) & 3.8 & 60 & 1.1$^{\ast}$ & $3.7\pm0.5$&\\
80& 060 & I &8(4$^b$) &n& $<87(91)$& 1.5 & 86 & 1.1$^{\ast}$ & - &\\
84& & 0 &3  &n& $<223$& 0.5 & 74 & 0.04 & - &\\
101&& I &1  &y& 227.8 & 0.2 & 463 & 1.0 & H&\\
Bolo62& 081 &0&2&n& $<98$ & 0.3 & 103 & 0.04 & H$^{ll}$&\\
- & 068 & 0 & 4 & n & $<90$ & -  & 64$^{\ast}$ & 0.45$^{\ast}$ & - &  \\
- & 080 & - & 10$^{\dagger}$(6$^b$) & y & 90.1 & -  & 34$^{\ast}$ & 0.64$^{\ast}$ & - &  \\
- & 090 & - & 3 & y & 114.0 & -  & 114$^{\ast}$& 0.57$^{\ast}$ & - &  \\
- & 092 & - & - & y & 206.7 & -  & 47$^{\ast}$ & 0.29$^{\ast}$ & - & \\
- & 109 & - & 17$^{\dagger}$(9$^b$) & n & $<125$ & - & 345$^{\ast}$& 0.26$^{\ast}$ & - & \\
\hline
\end{tabular}
\begin{minipage}{\textwidth}
{
Note: Where no outflow force is given an ``H'' indicates that an outflow has been detected by {\sc harp}, and ``H?'' indicates a \emph{possible} outflow (Hatchell et~al. 2007b) but no outflow force value is available, a suffix of H$^{ll}$ indicates the lower line-wing detection criterion (Hatchell \& Dunham 2010). 

$^a$ Improved fit to modified greybody with $\beta=1.5$.

$^b$ Improved fit to modified greybody with $\beta=2.0$.

$^c$ Unresolved from neighbouring source.

$^d$ Flux density confused by neighbouring source.

$^{\dagger}$ Greybody fitted using flux densities from Di~Francesco et~al. (2008).

$^{\ast}$ From DCE08.
}
\end{minipage}
\end{table*}
\begin{table*}
\caption{Physical parameters of protostellar sources outside Perseus. Columns are [1] source number from the Dunham et~al. (2008) catalogue; [2] association; [3] protostellar Class; [4] flux density predicted from vibrational dust emission at 16\,GHz; [5] flag on presence of radio counterpart; [6] 1.8\,cm radio flux density in excess of greybody emission measured by the AMI-LA; [7] envelope mass; [8] Bolometric temperature; [9] Bolometric luminosity from DCE08; [9] references for physical properties and class. \label{tab:nonp}}
\begin{tabular}{cccccccccc}
\hline \hline
[DCE08] & Other names & Class & S$_{\rm{pred,16}}$ & radio det. & $S_{16}$ & $M_{\rm{env}}$ & $L_{\rm{bol}}$ & $T_{\rm{bol}}$ & refs.\\
 &&&($\mu$Jy)&(y/n)&($\mu$Jy)&(M$_{\odot}$)&(L$_{\odot}$)&(K)&\\
\hline
005 & IRAS04381+2540 & I & 102 & y & 637.2 & $2.8\pm0.05$ & $0.67\pm0.02$ & 131 & K08; DiF08\\
044 & L1251-A IRS3   & O & 38 & y & 119.5 & $11.09\pm0.44$ & 0.8 & 24 & L10; W07\\
045 & L1251-A IRS4   & O & 4  & y & 224.5 & $16.46\pm0.68$ & 0.6 & 21 & L10; W07\\
048/049 & L1251-B IRS1/2 & O/I$^a$& 141 & y & 1135.7 & 17.2$^b$ & 10/0.6 & 87/140 & L06; DiF08; W07\\
\hline
043 & L1251-A IRS2   & I & 23 & n & $<125$ & $14.15\pm2.10$ & 0.4 & 230 & L10\\
\hline
\end{tabular}
\begin{minipage}{\textwidth}
References: Kauffmann et~al. 2008 (K08); Di~Francesco et~al. 2008 (DiF08); Lee et~al. 2010 (L10); Wu et~al. 2007 (W07); Lee et~al. 2006 (L06).\\
$^a$ as classified by this work, see text for details.\\
$^b$ as calculated by this work, from 850\,$\mu$m flux densities in Lee et~al. 2006.
\end{minipage}
\end{table*}
\begin{figure*}
\centerline{\includegraphics[width=0.4\textwidth]{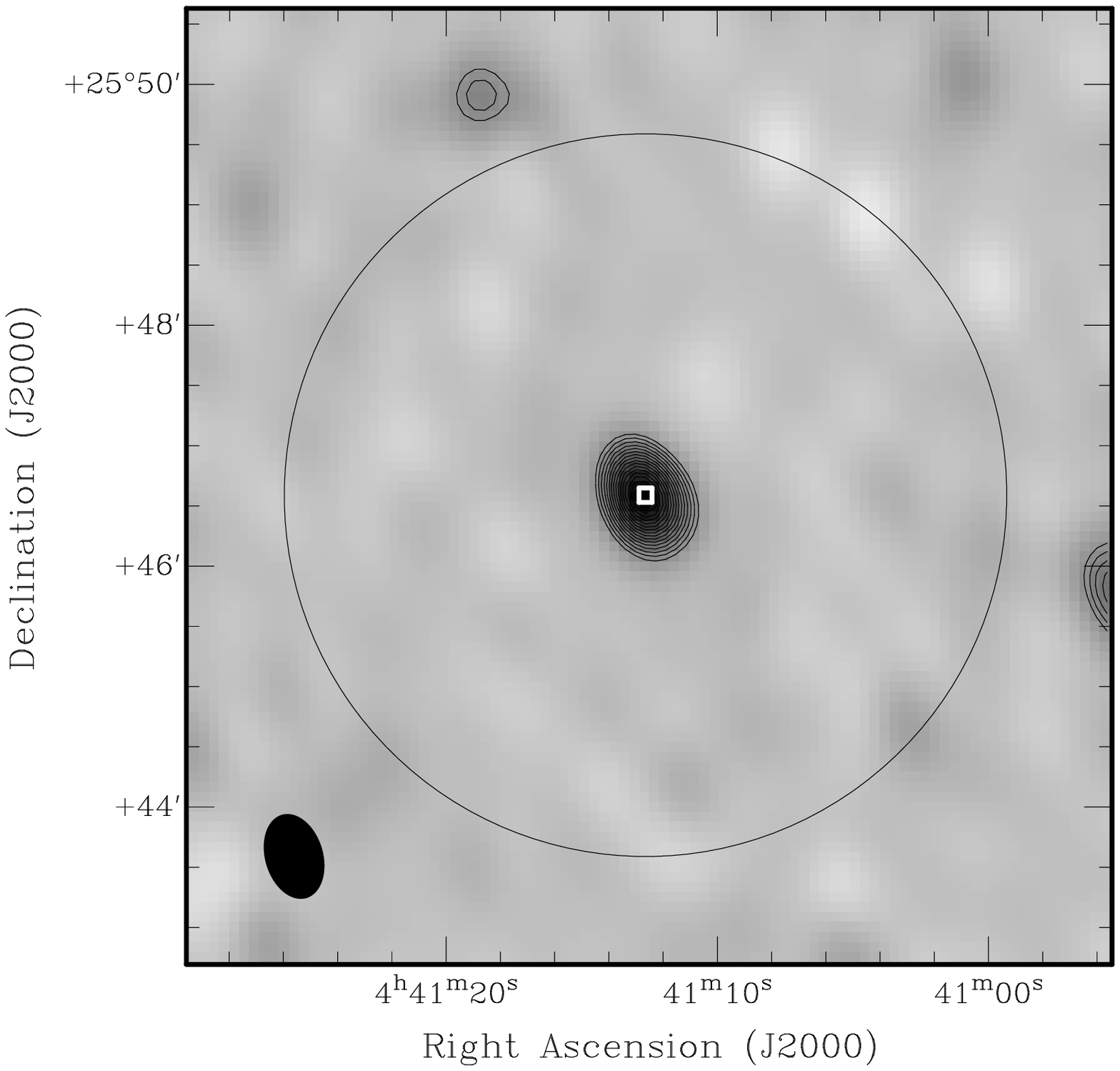}\qquad\includegraphics[width=0.4\textwidth]{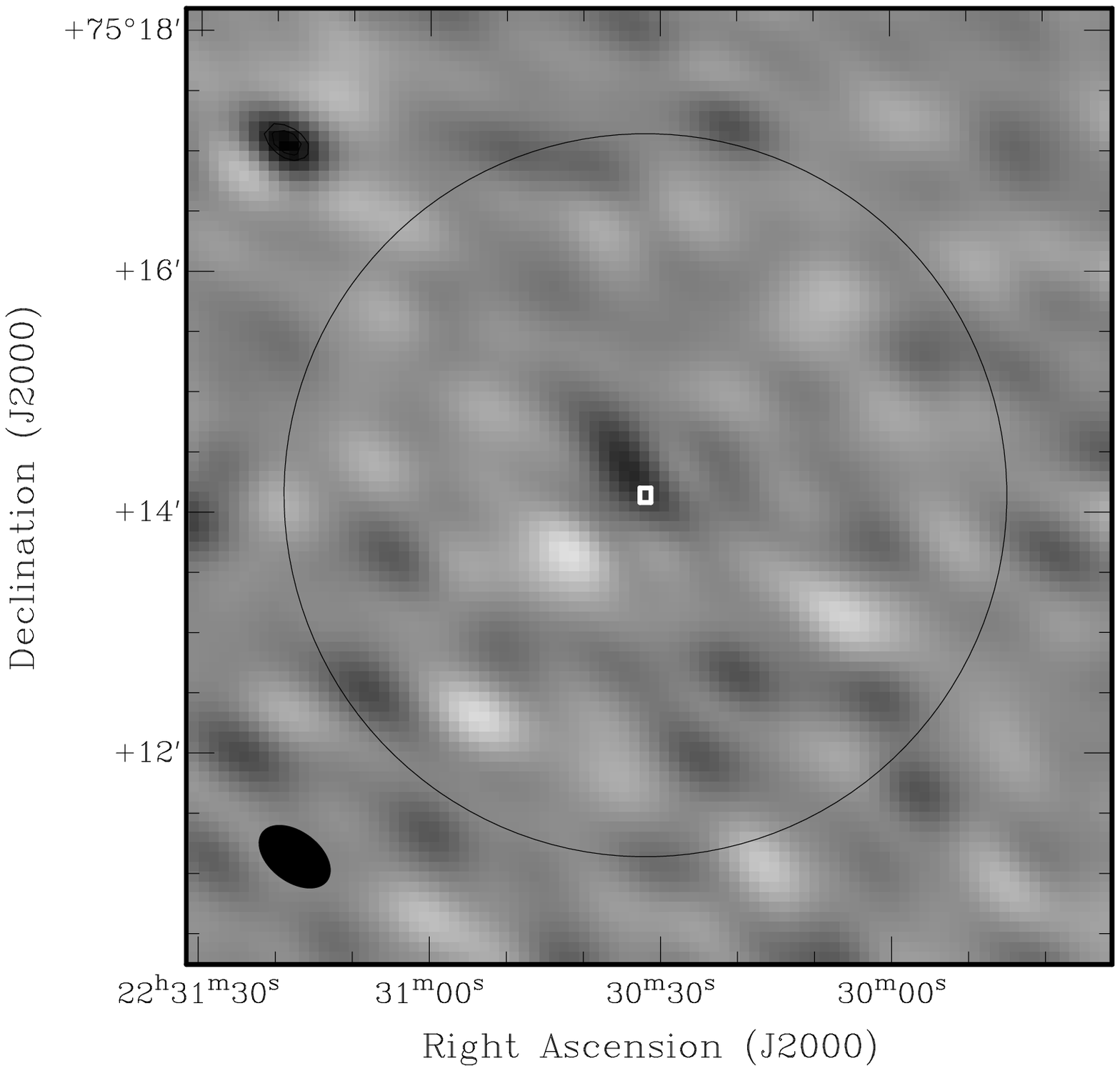}}
\centerline{DCE08-005\hspace{0.4\textwidth}DCE08-044}
\centerline{\includegraphics[width=0.4\textwidth]{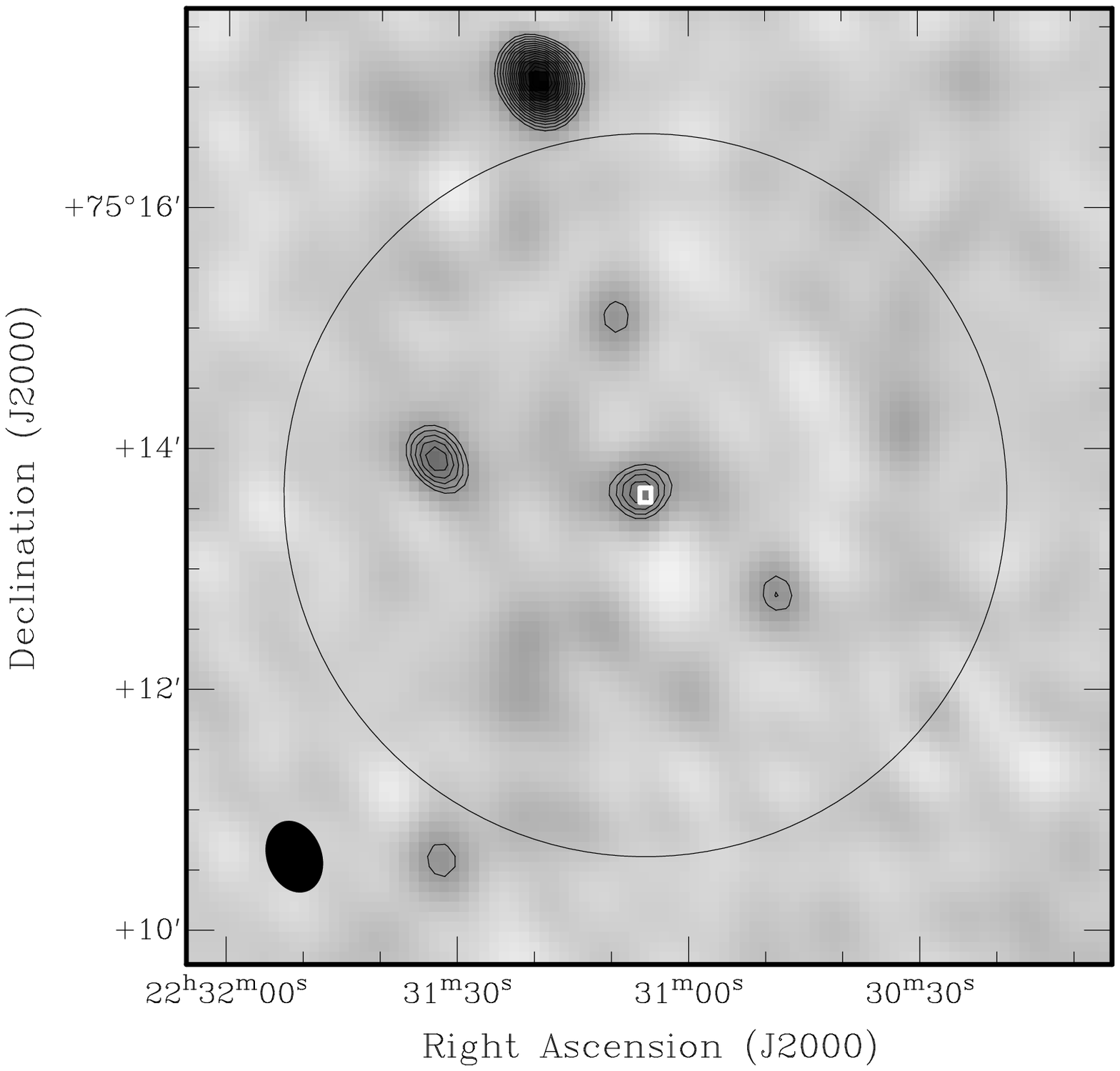}\qquad\includegraphics[width=0.4\textwidth]{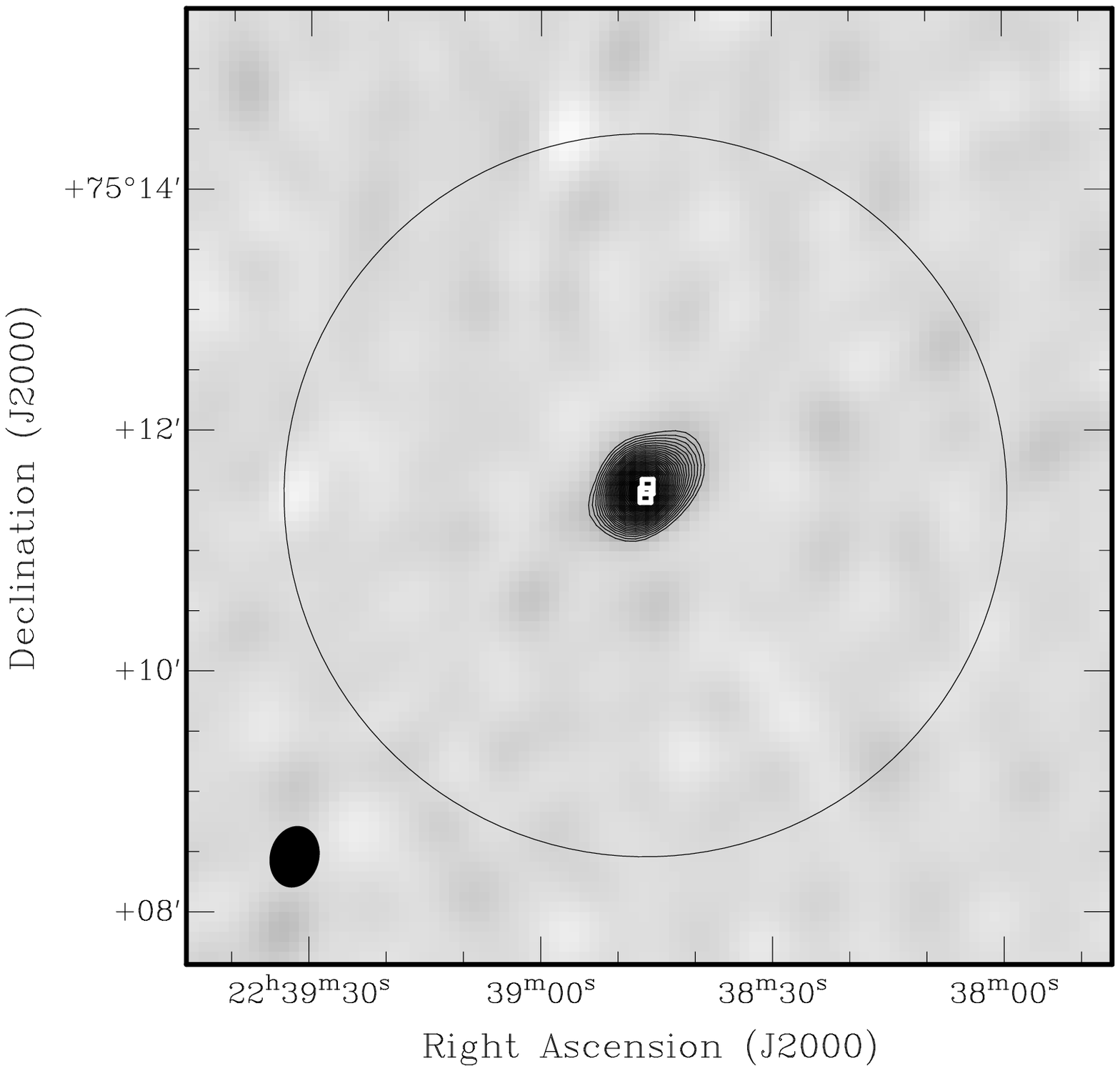}}
\centerline{DCE08-045\hspace{0.4\textwidth}DCE08-048/049}
\centerline{\includegraphics[width=0.4\textwidth]{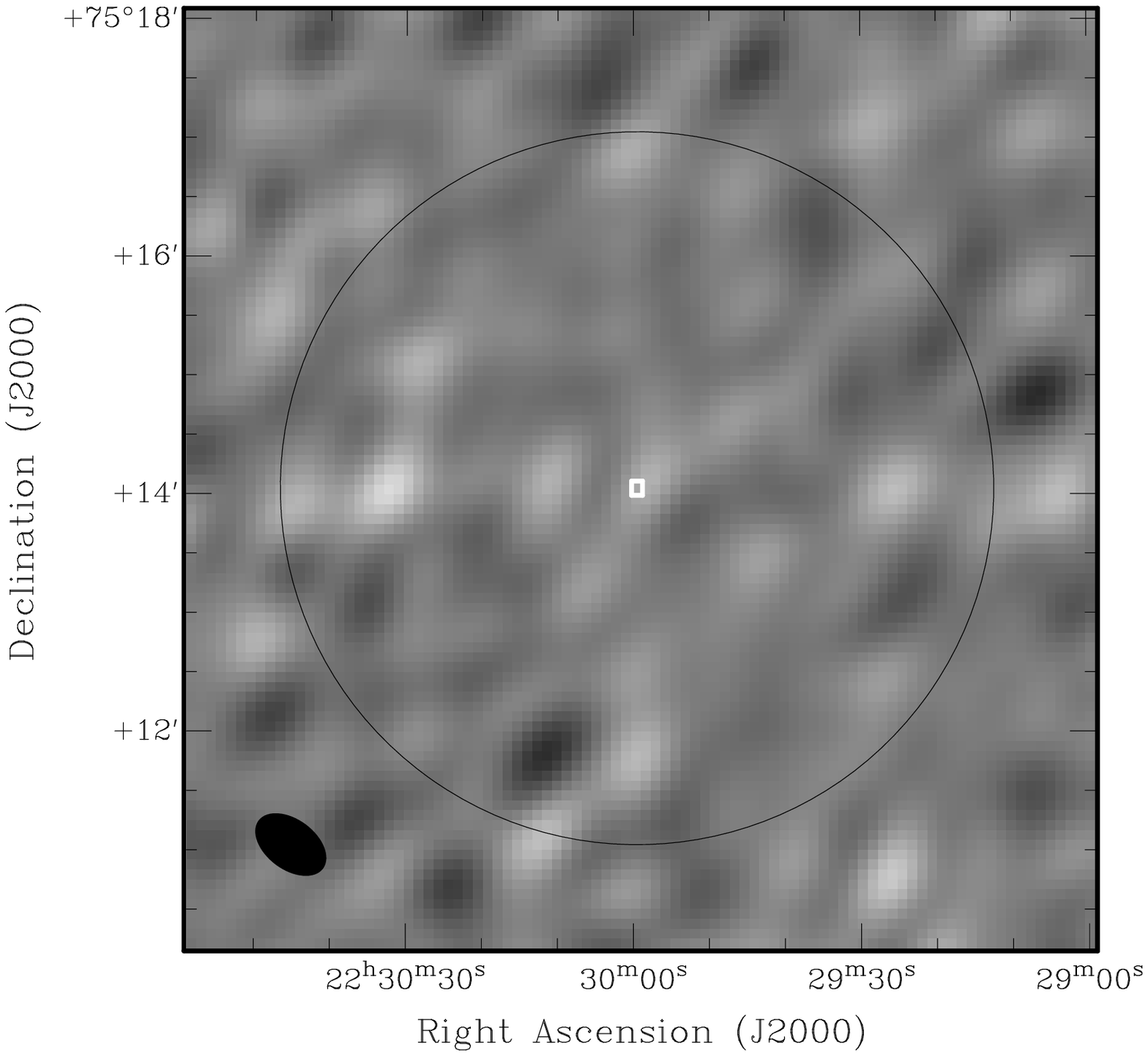}}
\centerline{DCE08-043}
\caption{Sources lying outside the Perseus region. Greyscale and contours at 16\,GHz from the AMI-LA are shown with no correction for the primary beam attenuation. Contours are shown at 5, 6, 7\,$\sigma_{\rm{rms}}$ etc., where values for $\sigma_{\rm{rms}}$ for each field can be found in Table~\ref{tab:obs}. The position of each source from Table~\ref{tab:nonp} is shown as an unfilled square. The FWHM of the AMI-LA primary beam is shown as a circle and the synthesised beam for each map as a filled ellipse in the bottom left corner. A full set of maps for the remainder of the sources observed in this work can be found in Appendix~\ref{app:maps}.\label{fig:none}}
\end{figure*}

\subsection{Expected contamination by extragalactic radio sources}

At 16\,GHz we expect a certain number of extragalactic radio sources to be seen within each of our fields. Following Paper I, to quantify this number we use the 15\,GHz source counts model from de Zotti et~al. (2005) scaled to the 10C survey source counts (AMI Consortium: Davies et~al. 2010).  The average rms noise from our datasets is $\simeq25\,\mu$Jy\,beam$^{-1}$ and from this model we predict that we should see 0.07 sources arcmin$^{-2}$, or $\simeq 2$ radio sources within a 6\,arcmin FWHM primary beam above a 5\,$\sigma$ flux density of 125\,$\mu$Jy. Within the Perseus region we would therefore expect to detect $\approx 26\pm5$ extragalactic radio sources, and in the remainder of the sample a further $\approx 9\pm3$ sources. Tables~\ref{tab:moresources1} \&~\ref{tab:moresources2} shows the full source list for objects detected at $>5\,\sigma$ in the fields. Making the assumption that all sources which cannot be identified with a previously known protostellar object are extragalactic, we find 21 radio sources in the Perseus fields, consistent with our prediction. In the fields outside the Perseus region we find only 3 non-protostellar sources. Although this is low compared to the prediction we suggest that this is simply because the covered sky area is small and that this is a statistical outlier, consistent at 2\,$\sigma$. Of these sources, seven are matched to known radio sources in the NVSS catalogue (Condon et~al. 1998) at 1.4\,GHz and six also have identifications in the WENSS catalogue at 327\,MHz (de~Bruyn et~al. 1999). Using this data we can identify all of these objects as steep spectrum radio sources with an average spectral index of $\alpha_{0.327}^{16}=-0.88\pm0.15$. Assuming a canonical spectral index for non-thermal emission of $\alpha=-0.7$, and taking the completeness limit of the NVSS survey as 3.4\,mJy (Condon et~al. 1998) we find that all but three of the remaining unmatched radio sources have integrated flux densities below this threshold when extrapolating from 16\,GHz. 

Of the three sources which do not lie below the NVSS detection threshold, one is extended and has a much lower peak flux density making it unlikely to have been detected in NVSS. The remaining two are AMI-09 and AMI-44; both sources have flux densities above 1\,mJy at 16\,GHz and are relatively compact. AMI-09 is identified as VLA-10 in Rodriguez et~al. (1999) who surveyed the HH~7-11 region at 3.5\,cm with the VLA. We also identify it with [EDJ09]-176 from Evans et~al. (2009) who classify it as a YSO, and [GMM08]-3 from Gutermuth et~al. (2008) who identify this YSO as an embedded Class I source. AMI-44 is also tentatively identified as a Class I protostar in the IC\,348 nebula from the \emph{Spitzer} catalogue of Muench et~al. (2007).

\section{Protostellar Sources}
\label{sec:pstars}

\subsection{Perseus Region}

Those fields which lie within the Perseus region surveyed in the sub-mm by H07 contain a number of known protostellar sources in addition to those selected from the \emph{Spitzer} catalogue of DCE08. Table~\ref{tab:phys} lists all the protostellar sources within these fields from H07 and identifies them with sources detected in the AMI-LA catalogue in this work. Where a known object is not detected by the AMI-LA an upper limit on the 1.8\,cm radio flux density of that source is given. These limits take into account the position of the source within the AMI-LA primary beam. Since source detection is performed in the maps before correction for the  primary beam this means that the true limit on the unattenuated flux density will be given by $S_{\rm{lim}}<5\,\sigma_{\rm{th}}/A({r})$, where $A({r})$ is the primary beam attenuation at a radial distance ${r}$ from the pointing centre. 

A number of the physical properties listed in Table~\ref{tab:phys} including the radio luminosity depend on the distance to Perseus. Although many studies assume a distance for Perseus of 320\,kpc (e.g. H07), we here assume a distance of $D=250$\,kpc in order to remain consistent with DCE08. Where necessary, physical quantities from the literature have been corrected accordingly.

\subsection{Additional targets}

There are six targets in our sample which lie outside the Perseus region. The first of these objects, [DCE08]-005, lies within the Taurus molecular cloud and is also known as IRAS04381+2540. We use the physical parameters derived for this core by Kauffmann et~al. (2008) from MAMBO 350\,$\mu$m data. The five remaining targets are all found within the L1251 dark cloud. [DCE08]-043, 044 and 045 are found in the vicinity of Core A, and [DCE08]-048 and 049 in the vicinity of Core B. Their precise identifications are listed in Table~\ref{tab:nonp}. [DCE08]-048 and [DCE08]-049 are unresolved by the AMI-LA and are therefore listed together. Lee et~al. (2006) found that the these sources were both borderline Class 0/I based on their bolometric temperatures and the ratio of the sub-mm to bolometric luminosity. By also considering the ratio of 3.6\,$\mu$m to 850\,$\mu$m flux density we classify [DCE08]-048, identified with L1251-B IRS1, as Class 0; and [DCE08]-049, identified with L1251-B IRS2, as Class I. The physical parameters and classifications of those objects in L1251-A are taken from Lee et~al. (2010).

\subsection{Expected vibrational dust contribution}
\label{sec:dust}

For each of the sources listed in Table~\ref{tab:phys} we constrain the contribution of thermal dust emission to the measured flux density at 16\,GHz by fitting to sub-mm data at 1.1\,mm (Enoch et~al. 2004), and 850 and 450\,$\mu$m (Hatchell et~al. 2007). For the sources in Table~\ref{tab:nonp} we use sub-mm data from the references in that Table. Following Paper I we fit modified greybody spectra of the form
\begin{equation}
S_{\nu} = \nu^{\beta}B_{\nu}(T_{\rm{d}}),
\end{equation}
where $\nu$ is the frequency of the data point, $\beta$ is the opacity index for which we assume a canonical value of $\beta=1.8$, $B_{\nu}$ is the Planck function for a dust temperature $T_{\rm{d}}$. A value of $\beta=1.8$ agrees with that used by Hatchell et~al. (2007) and is close to the theoretically derived value of $\beta_{\rm{OH5}}=1.85$ found for coagulated grains with thin ice mantles between 350\,$\mu$m and 1.3\,mm (Ossenkopf \& Henning 1994). 

Tables~\ref{tab:phys}~\&~\ref{tab:nonp} list the predicted flux density due to thermal dust emission at 16\,GHz from these fits. Where a value of $\beta=1.8$ does not provide a good fit to the data we repeat these fits using alternate values of $\beta=1.0, 1.5\,\,\&\,\,2.0$. Where appropriate a second prediction for the 16\,GHz flux density due to thermal dust and based on the minimum $\chi^2$ value of $\beta$ is also shown in Table~\ref{tab:phys}. In practice a value of $\beta=1.0$ is not preferred for any of our sources. This is perhaps unsurprising as the sample are predominantly young protostars with low values of $T_{\rm{d}}$, and as such are expected to have larger values of $\beta$ in general (Planck Collaboration: Douspis et~al. 2011). For those objects which are only detected at one of the three sub-mm wavelengths we simply extrapolate to 16\,GHz using the canonical $\beta$. We note that there is a degree of uncertainty in these predictions, largely due to the systematic uncertainties in sub-mm flux densities, which also in general limit the determination of $\beta$ (Shirley et~al. 2011a). In addition, a single value of $\beta$ may not hold to cm-wave frequencies, notably in the case of objects with disks (Shirley et~al. 2011a), but we consider this model to be a reasonable approximation in the absence of additional sub-mm and mm-wave data.

In what follows we use the measured flux density at 1.8\,cm with the predicted greybody contribution \emph{subtracted} to calculate the radio luminosity. This is to ensure that the values used are representative only of the radio emission and do not include contributions from the thermal dust tail which varies greatly between sources (see Tables~\ref{tab:phys} and \ref{tab:nonp}) and which might therefore influence any conclusions being drawn from the distributions examined in the later stages of this paper. 

The average ratio of predicted thermal dust emission to residual radio emission is approximately 14\%. This shows that although the contribution is small it is not negligible. In addition we can see that the ratio is lower for detected Class I sources (9\%) compared with Class 0 sources (18\%). This might be expected due to their comparatively smaller dust envelopes, but could also indicate an increase in radio emission relative to thermal dust emission for these objects.

\subsection{Luminosity and Outflow Force Correlations}
\label{sec:lcorr}

\begin{figure}
\centerline{\includegraphics[angle=-90,width=0.5\textwidth]{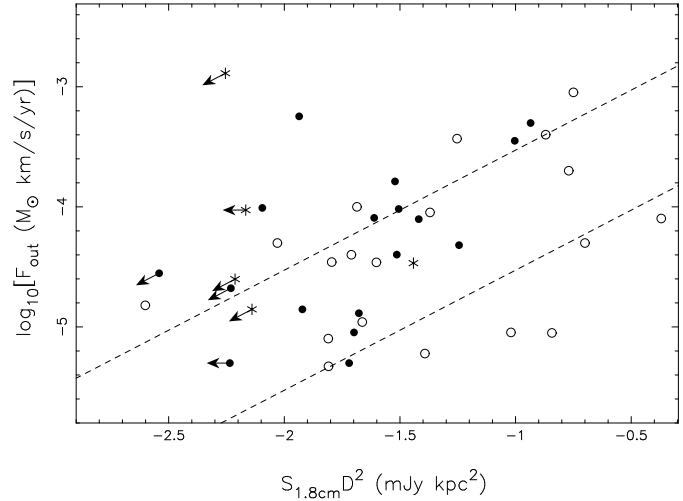}}
\caption{Correlation of 1.8\,cm radio luminosity with outflow force. Data from Paper I are shown as unfilled circles, Class 0 objects  from this work are shown as filled circles and Class I objects as stars. The dashed lines show the theoretical relationship between 1.8\,cm radio luminosity and outflow force (Curiel et~al. 1989) for an efficiency of $\eta=1$ (minimum required force; lower line) and an efficiency of $\eta=0.1$ (upper line). \label{fig:fcorr}}
\end{figure}

\begin{figure}
\centerline{\includegraphics[angle=-90,width=0.5\textwidth]{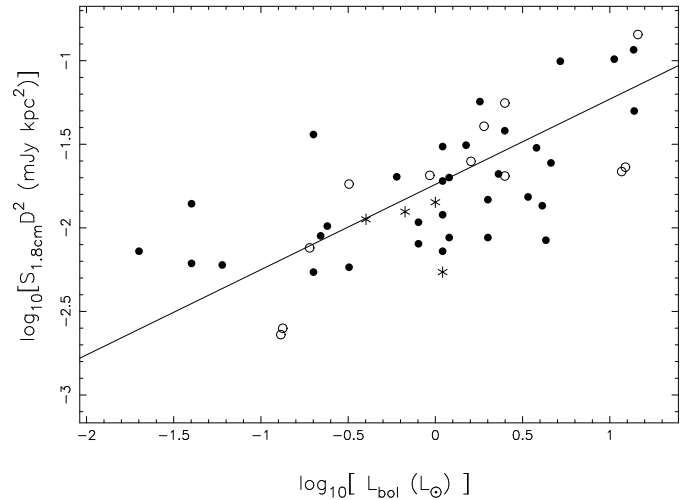}}
\vskip .1in

\caption{Correlation of radio luminosity with bolometric luminosity for objects detected at $>5\,\sigma$. Class 0 sources from this work are shown as filled circles, Class I sources from this work are shown as stars, and sources from Paper I are shown as unfilled circles. The best fitting correlation is shown as a solid line. \label{fig:lcorr1}}
\end{figure}

\begin{figure}

\centerline{\includegraphics[angle=-90,width=0.5\textwidth]{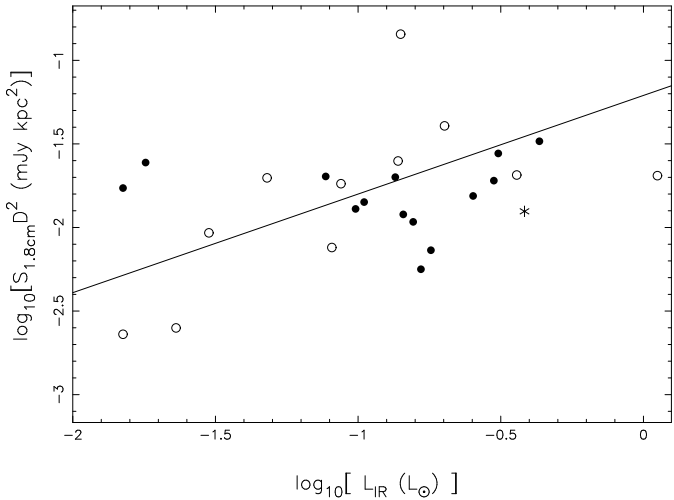}}
\vskip .1in

\caption{Correlation of radio luminosity with infra-red luminosity for objects detected at $>5\,\sigma$. Class 0 sources from this work are shown as filled circles, Class I sources from this work are shown as stars, and sources from Paper I are shown as unfilled circles. The best fitting correlation is shown as a solid line. \label{fig:lcorr2}}
\end{figure}

A detailed investigation of the possible causes for a correlation of the radio luminosity with the bolometric luminosity, infrared luminosity and outflow force was presented in Paper I and references therein. We do not revisit these discussions here but utilise the combined  data from this work and Paper I to redraw these correlations. Values for the physical characteristics of these sources are listed in Table~\ref{tab:phys}. In the case of outflow force the letter `H' indicates that an outflow has been detected with {\sc harp}, but that no value for its momentum flux is available in the literature. The correlations between these parameters and the radio luminosity are shown in Figs.~\ref{fig:fcorr}, \ref{fig:lcorr1} and \ref{fig:lcorr2}. 

The correlation of radio luminosity with outflow force is shown in Fig.~\ref{fig:fcorr}. Although it is evident that no improvement in this correlation is made by the addition of these new measurements, this data is consistent with the loose trend measured in Paper I. As discussed at length in Paper I the lack of a well-defined correlation in this case may be due to the errors inherent in measurements of outflow force. 

Although Class I objects are generally considered to have weaker outflows than Class 0 objects there is no clear evolutionary division in the sources sampled here. Indeed the outflow force values for the Class I objects in this sample span almost as broad a range as those from Class 0. As in Paper I we plot the minimum outflow force required to explain the cm-wave emission as arising from shock ionisation (Anglada 1995); see Fig.~\ref{fig:fcorr}. Although the data seems to follow the general trend described by this model not all objects lie above the limit indicated for an efficiency of $\eta=1$, indicating that their radio emission cannot arise, at least solely, from shock ionisation. 

The correlation of radio luminosity with bolometric luminosity is shown in Fig.~\ref{fig:lcorr1}. All sources for which a bolometric luminosity exists in the literature are included, see Tables~\ref{tab:phys} and \ref{tab:nonp} for values and references. The data from this work are very consistent with those from Paper I.
The improved correlation combining the data from this paper with that of Paper I, shown in Fig.~\ref{fig:lcorr1}, is
\begin{eqnarray}
\nonumber \log[L_{\rm{1.8\,cm}} (\rm{mJy\,kpc}^2)] & = & -(1.74\pm0.18)\\
&&+(0.51\pm0.26)\log[L_{\rm{bol}} (\rm{L}_{\odot})].
\label{equ:corrl}
\end{eqnarray}

The correlation of radio luminosity with infra-red luminosity is shown in Fig.~\ref{fig:lcorr2}. The data shown are those for which a source is found in DCE08, and which have an IR luminosity derived in that work. The correlation with IR luminosity found in Paper I, and shown as a solid line in Fig.~\ref{fig:lcorr2}, is a good fit to the combined data sets. Notable in this plot are the possible VeLLO sources [DCE08]-065 and [DCE08]-073, which both have radio luminosities in excess of the general trend. A further object which does not follow the correlation is [DCE08]-055, however there is an obvious explanation for this behaviour. Although we associate [DCE08]-055 with [H07]-028 due to proximity, the radio detection is in fact also completely unresolved from [H07]-027. Since there is no infra-red luminosity available in the literature for [H07]-027 we describe the $L_{\rm{IR}}$ of [DCE08]-055 ([H07]-028) as being a lower limit, and exclude it from the fitted data shown in Fig.~\ref{fig:lcorr2}.

The improved correlation combining the data from this paper with that of Paper I, is
\begin{eqnarray}
\nonumber \log[L_{\rm{1.8\,cm}} (\rm{mJy\,kpc}^2)] & = & -(1.23\pm0.39)\\
&&+(0.59\pm0.34)\log[L_{\rm{IR}} (\rm{L}_{\odot})].
\label{equ:corrIR}
\end{eqnarray}

\subsection{Correlation with Envelope Mass}

For each of the objects cross-identified with the 850\,$\mu$m Scuba catalogue of H07 we can also investigate the correlation of the radio luminosity with core envelope mass. These masses were determined using the 850\,$\mu$m flux density in the following way:
\begin{equation}
M=\frac{S_{\nu}D^2}{\kappa_{\nu}B_{\nu}(T_{\rm{d}})},
\end{equation}
making the assumptions that the dust temperature, $T_{\rm{d}}=10$\,K and the dust opacity, $\kappa_{850\mu\rm{m}}=0.0012$\,m$^2$\,kg$^{-1}$. The opacity is known to follow the distribution $\kappa_{\nu}=\kappa_0(\nu/\nu_0)^{\beta}$, where $\beta=2$. 

\begin{figure}
\centerline{\includegraphics[angle=-90,width=0.5\textwidth]{./PER-FIG/hatchell_menv_corr.ps}}
\vskip .1in
\centerline{(a)}

\centerline{\includegraphics[angle=-90,width=0.5\textwidth]{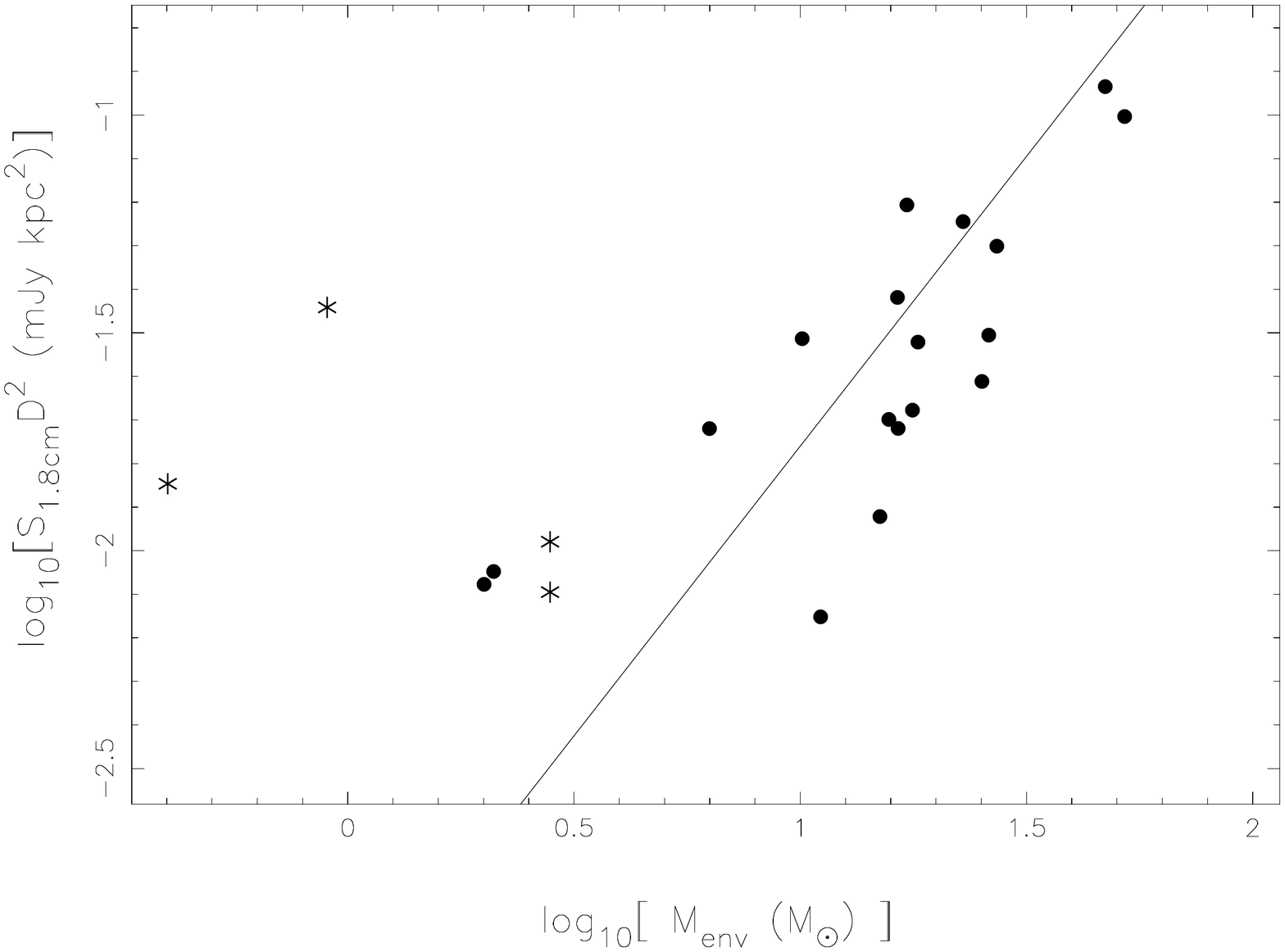}}
\vskip .1in
\centerline{(b)}

\caption{Correlation of radio luminosity with envelope mass.  (a) Linear correlation, and (b) Logarithmic correlation. In (a) two data points are presented for each object: unfilled circles show the sub-mm greybody predictions for 16\,GHz luminosities from thermal dust emission, filled circles and stars show the measured radio luminosities in excess of the greybody predictions for Class 0 and Class I objects, respectively; a 16\,GHz luminosity prediction from envelope mass is shown as a dashed line; in (b) the radio luminosity data only is plotted; a best fitting regression is shown as a solid line, the relationship $L_{\rm{rad}} \propto {M}^{4/3}$ is shown scaled to fit the Class 0 data as a dashed line, and the predicted relationship based on the that of radio and bolometric luminosity is shown as a dotted line, see text for details. \label{fig:corr2}}
\end{figure}

Reversing this relationship we can see that luminosity ($S_{\nu}D^2$) is a linear function of core envelope mass, $M$. Using the envelope masses derived from the 850\,$\mu$m measurements and the known behaviour of $\kappa_{\nu}$ we can predict the radio luminosity from the dust mass at 1.8\,cm. This prediction is shown as a dashed line in Fig.~\ref{fig:corr2}. The predicted luminosities for each object at 16\,GHz based on a greybody fitted to sub-mm data, the flux densities for which are listed in Column [4] of Tables~\ref{tab:phys} and \ref{tab:nonp}, are shown as unfilled squares in this plot and the radio luminosities in excess of these predictions, i.e. the difference between the total radio luminosity and the greybody prediction, are shown as filled circles and stars for Class 0 and Class I objects, respectively. The sub-mm greybody predictions for the sample are well fitted by the mass prediction model, with perhaps some evidence for a lower average value of $\kappa_{\nu}$, which is known to vary, or alternatively a lower value of the emissivity index $\beta$ used to calculate the greybody emission. For a detailed comparison of current measured opacities for low mass cores we refer the reader to Shirley et~al. (2011b). The value of $\kappa_{850\mu\rm{m}}$ used here agrees with that derived by Shirley et~al. (2011b) for the low mass Class 0 core B335 who found a range of values $\kappa_{850\mu\rm{m}}=(1.18-1.77)^{+0.36}_{-0.24}\times 10^{-3}$\,m$^2$\,kg$^{-1}$. 

It is clear from Fig.~\ref{fig:corr2} that a correlation also exists between envelope mass and radio luminosity, with a Pearson correlation coefficient of $r=0.89$ indicating a strong positive correlation. However, this correlation does not seem to hold as strongly, if at all, for the detected Class I objects in this sample which are all found at the low-mass end of the relationship and deviate from the trend. Unlike the luminosity and outflow force correlations this trend seems to show a dependency on protostellar evolution, with all Class I objects lying at the low mass end. The fact that the Class I objects also deviate from the general correlation seen in the Class 0 sources suggests, similarly to the  detection statistics (see \S~\ref{sec:stats}), that the radio emission from Class I objects is heavily influenced by environmental factors rather than arising as a consequence of an intrinsic mechanism. 

The correlation fitted to the Class 0 sources in this sample is
\begin{eqnarray}
\nonumber \log[L_{\rm{1.8\,cm}} (\rm{mJy\,kpc}^2)] & = & -(2.23\pm0.65)\\
&&+(0.68\pm0.62)\log[M_{\rm{env}} (\rm{M}_{\odot})],
\label{equ:corrm}
\end{eqnarray}
and is plotted as a solid line in Fig.~\ref{fig:corr2}.

It is known that a correlation exists between bolometric luminosity and core mass (see e.g. Planck collaboration: Montier et~al. 2011) with the form $\log{L_{\rm{bol}}} = \log{A}+0.67\log{M_{\rm{env}}}$, where the constant $A$ depends on the surface density. Combining this relationship with that fitted between radio and bolometric luminosity in this work, Eq.~\ref{equ:corrl}, we can predict a weak correlation between radio luminosity and core mass with the form $\log{L_{\rm{rad}}} = \log{A'}+0.31\log{M_{\rm{env}}}$. The relationship shown in Equ.~\ref{equ:corrm} is broadly consistent with this correlation due to the large scatter in the data, and is shown in Fig.~\ref{fig:corr2}(b) as a dotted line. However, it is unclear why this relationship should fail to apply in the case of Class I objects. It can be seen from Fig.~\ref{fig:lcorr1} that the bolometric luminosities of the Class I sources in this sample show no deviation from the general trend, nor with any evolutionary distinction evident in their distribution. 

As sources evolve through the Class~0 and Class~I stages, both the envelope mass and the average mass loss rate in the outflow decrease (Bontemps et~al. 1996; Ladd \& Fuller 2002; Arce \& Sargent 2006). The observed correlation of radio luminosity, $L_{\rm{rad}}$, with envelope mass could also be seen as a correlation with mass loss rate. 

Models of radio emission from both ionised spherical stellar winds (Panagia \& Felli 1975) and collimated winds (Reynolds 1986) both predict a correlation between radio luminosity, $L_{\rm{rad}}$, and the rate of stellar mass loss, $\dot{M}$. In the canonical spherical case with $n_{\rm{e}}\propto r^{-2}$
\begin{eqnarray}
\nonumber \left[\frac{S_{\nu}d^2}{\rm{mJy\,kpc^2}}\right] &= &5.12
 \left[\frac{\nu}{10\,\rm{GHz}}\right]^{0.6}
 \left[\frac{T_{\rm{e}}}{10^4\,\rm{K}}\right]^{0.1}
 \left[\frac{\mu}{1.2}\right]^{-4/3} \\
 &&\left[\frac{v_{\ast}}{10^3\,\rm{km\,s^{-1}}}\right]^{-4/3}
 \bar{Z}^{-2/3}
 \left[\frac{\dot{M}}{10^{-5}\,\rm{M}_{\odot}\,\rm{yr}^{-1}}\right]^{4/3}
\label{equ:pf75}
\end{eqnarray}
(Panagia \& Felli 1975). For the case of a collimated outflow Reynolds (1986) also predicted the relationship
\begin{equation}
L_{\rm{rad}} \propto \dot{M}^{4/3},
\end{equation}
where the constant of proportionality included weak dependencies on ionisation fraction, collimation, temperature and outflow inclination. Although the best-fit regression to the Class 0 data shown in Fig.~\ref{fig:corr2}(a) \& (b) is not exactly equal to this value, it can be seen that these data are not inconsistent with the predicted relationship, plotted as a dashed line in Fig.~\ref{fig:corr2}(b). 

In recent models, mass accretion is suggested to be episodic, raising luminosities and temperatures and switching protostars back and forth between Class~0 and Class~I (Dunham et~al. 2010). If the variable accretion model were correct, and radio luminosities did correlate with mass loss rate in the outflow, then Class~I sources should have radio luminosities higher by an order of magnitude than Class~0 sources. We find that the small number of Class~I sources which are detected in this sample have much higher radio luminosities relative to the correlation seen with envelope mass for Class~0. However, the high number of non-detections for Class~I sources also suggests that mass loss rate is not the only factor controlling the radio luminosity for these objects.

We also note that the observed correlation could also be explained by a population of dust with a similar mass but a higher opacity, $\kappa_{850\,\mu\rm{m}}\approx 0.006$\,m$^2$\,kg$^{-1}$. This value is very high compared with the range of values collected by Shirley et~al. (2011b) and, even considering the degenerate contributions of mass, opacity and temperature to this relationship, it would seem unlikely that the emission we see at centimetre wavelengths is not in fact free--free emission but thermal emission from a separate population of dust. Alternatively the emission could also arise as a consequence of rotational emission from a population of very small dust grains. If these VSGs were mixed through the dense dust core as a function of density then the amount of rotational emission would also be a function of dust mass. However, without additional supporting evidence we do not explore this possibility further.

\subsection{Correlation with Bolometric Temperature}
\label{sec:tbol}

The bolometric temperature, values taken from H07, is representative of the dust temperature in the cores - rather than the temperature of any ionized gas. Consequently the lack of correlation with radio luminosity is perhaps not surprising.

On average the bolometric temperature is thought to increase with bolometric luminosity, however within individual classes of protostellar object this trend is extremely weak and indeed it is possible that it is itself simply an effect of observational biases in the sub-mm (Kauffmann et~al. 2008). Although MAMBO follow-up of the \emph{Spitzer} c2d catalogue showed a correlation between bolometric temperature and envelope mass (Kauffmann et~al. 2008), this was only really evident when combining objects in Classes O, I and II. Individual classes did not portray a strong trend between these parameters and the same is true for the sample of objects considered here. When Classes O and I are considered together a weak negative correlation is found, $r=-0.41$; however if the Class 0 objects alone are considered then this correlation reduces even further, and in fact reverses, to $r=0.24$.

When comparing bolometric temperatures determined using IRAC data (H07) and those found from fluxes calculated using \emph{Spitzer} SEDs (DCE08) there is a marked difference for this sample, with those derived from IRAS being, on average, a factor of 1.6 higher than the corresponding \emph{Spitzer} values. Bolometric temperature is one of the indicators for protostellar class and within this sample three sources with $70\leq T_{\rm{bol}} \leq 150$, will have this indicator change from Class I to 0 when the \emph{Spitzer} temperature is used. For two of these sources, [DCE08]-060 ([H07]-80) and [DCE08]-064 ([H07]-074), their overall classification will change from Class I to Class 0 when \emph{Spitzer} temperatures are utilised rather than those from IRAC. However, neither is detected at 1.8\,cm by the AMI-LA. The third, [DCE08]-088 ([H07]-076), is already identified as Class 0 and is detected.

\subsection{Detections and Non-detections}
\label{sec:stats}

From the 28 objects targeted in the \emph{Spitzer} catalogue of embedded sources (DCE08) there are nine in the Perseus region which we do not detect. Of these, four are designated Group 1 by DCE08 and are therefore most likely to be true embedded protostars. From these Group 1 objects two are Class I following the revised classifications from \S~\ref{sec:tbol} ([DCE08]-063 \& 064) and two are Class 0 ([DCE08]-068 \& 081). [DCE08]-068 is identified as Class 0 following an identification with [H07]-75 in Hatchell \& Dunham (2010), however we note that the positions of these sources are in fact offset by $45''$ making their correspondence uncertain. This may be an example of a false protostellar identification due to confusion with an external galaxy in DCE08. [DCE08]-081 was initially identified as Class 0 by H07, but no outflow was detected by Hatchell et~al. (2007b). An outflow was identified by Hatchell \& Dunham (2010) using the ``lower line-wing criteria", with the caveat that it may have been confused with the outflow from [H07]-081. Five of the six Group 3 objects are not detected. Of these two are considered starless ([DCE08]-107 \& 108), one is Class I ([DCE08]-060), one has no classification and no counterpart in H07 ([DCE08]-109). The remaining non-detection is [DCE08]-104, which is a Class 0 source. 

Of the sources outside the Perseus region all Group 1 objects are detected, although we note that [DCE08]-048 and 049 are unresolved and it is likely that the radio detection is dominated by [DCE08]-048. [DCE08]-043 (Class I), the only Group 3 object outside Perseus is not detected. From these numbers the radio detection rates are 81\,\% and 29\,\% for the Group 1 and Group 3 objects, respectively.

The detection statistics for the whole sample divided by protostellar Class and candidate Group are summarised in Table~\ref{tab:detstat}. The results are very similar to those found in Paper I for VeLLO and Class 0 sources. For Class I sources the detection rate is lower in this work compared with that of Paper I (71\,\%). The consistency in detection for Class 0 sources, and lack of consistency for Class I sources, may indicate that the radio emission mechanisms for the two types of protostellar objects differ. Consistency indicates that the emission from Class 0 objects is dependent on an intrinsic mechanism, whereas for Class I objects environmental factors are important leading to different detection statistics in different regions. We note that a comparison of the statistics for VeLLO sources is not conclusive due to the small number of these objects in this work. We also tabulate the detection statistics for the much larger number of protostellar sources which lie within our fields identified from the catalogue of H07, however we note that the primary beam attenuation away from the pointing centre will bias these detections non-uniformly. We include starless cores identified in H07 in these statistics, but do not list the positions for these objects in Table~\ref{tab:phys} as they are uniformly not detected. In spite of the differences between these samples the detection statistics for Class 0 and Class I objects are very similar for the original DCE08 sample and the extended dataset. We note that these numbers include [DCE08]-049 as a detection, although it is not possible to say conclusively whether it is contributing to the flux density of AMI-49, which combines [DCE08]-048 and [DCE08]-049. The effect of considering it a non-detection would be to reduce the Class I detection rate to 17(25)\% for the original sample and 18(22)\% for the extended sample. 

Since the correlation of radio luminosity with bolometric luminosity is the most well defined we examine the distribution of these non-detections with respect to this relationship. There are two distinct regions of non-detection. The first comprises low- and very low bolometric luminosity objects. The non-detection limits for these sources lie entirely above the general correlation and we can therefore assume that the fact they are not detected is biased by their low intrinsic luminosities. The second group is found at higher bolometric luminosities ($L_{\rm{bol}}>1$\,L$_{\odot}$) and lies predominantly below the general trend. For these objects there are two possibilities, firstly that their non-detection is simply an observational bias caused by the sensitivity limit of the observations, or secondly that they do not follow the general trend and have abnormally low, or zero, radio emission. There is no evident common characteristic between these sources which might be used to distinguish them from the general population, and in the absence of more sensitive observations it is not possible to say definitively if this is indeed the case as the sensitivity limit is not significantly removed from the general trend.

\subsection{VeLLOs}

There are three sources in this sample which can be defined as VeLLOs based on their modelled internal luminosities by DCE08, they are [DCE08]-064, 065 and 081. The source [DCE08]-073 has an infra-red luminosity of $L_{\rm{IR}}=0.018$\,L$_{\odot}$, well below the rough cut-off of $L_{\rm{IR}}=0.05$\,L$_{\odot}$ that generally indicates VeLLO status, which may also make it a VeLLO candidate or borderline case. Of these sources, we detect [DCE08]-065 and [DCE08]-073 at 1.8\,cm with the AMI-LA. Both of the detected sources are identified as Class 0 from H07. Of the undetected sources, [DCE08]-081 is identified as Class 0 and [DCE08]-064 as Class I by H07, although we note that [DCE08]-064 would be re-classified as Class 0 using its bolometric temperature from DCE08 as discussed in Section~\ref{sec:tbol}. [DCE08]-073 and [DCE08]-065 deviate from the general correlation trend of radio luminosity to infra-red luminosity, having a radio luminosities in excess of that expected, and this has already been noted in Section~\ref{sec:lcorr}. However, with two sources of similar infrared luminosity from Paper I having radio luminosities below the general trend it is likely that this is simply due to scatter in an under-sampled luminosity range rather than representing a characteristic deviation. [DCE08]-073 is the only VeLLO object with a measured outflow force and we note from Fig.~\ref{fig:fcorr} that the radio luminosity for this source is consistent with having arisen as a consequence of shock ionisation, unlike the VeLLO source L1014 (Shirley et~al. 2006) which has an outflow momentum significantly too low to explain its radio emission.

\begin{table}
\begin{center}
\caption{Summary of detection statistics for Table~\ref{tab:phys}. The original sample includes only objects from DCE08, listed in Table~\ref{tab:sample}. The extended sample includes all sources identified in H07 and listed in Table~\ref{tab:phys}, inclusive of the original sample. \label{tab:detstat}}
\begin{tabular}{cccc}
\hline\hline
Class & Present & Detected & \% \\
\hline
\multicolumn{4}{l}{\emph{Original sample}:}\\
VeLLO & 4 & 2 & 50 \\
O & 16(18) & 13(13) & 81(72) \\
I & 6(4) & 2(2)  & 33(50)$^{\dagger}$ \\
Starless & 2 & 0  & 0 \\
Group 1 & 21 & 17 & 81\\
Group 3 & 7 & 2 & 29\\
\multicolumn{4}{l}{\emph{Extended sample}:}\\
0 & 25(27) & 18(18) & 72(67) \\
I & 11(9) & 3(3) & 27(33)$^{\dagger}$ \\
Starless & 15 & 0 & 0\\
\hline
\end{tabular}
\begin{minipage}{0.4\textwidth}
Note: Figures in brackets indicate revised numbers following re-classification of sources based on bolometric temperatures derived from \emph{Spitzer} data in DCE08.\\
$^{\dagger}$ By considering [DCE08]-049 as a non-detection these figures are reduced to 17(25)\% for the original sample and 18(22)\% for the extended sample, see text for details.
\end{minipage}
\end{center}
\end{table}

\section{Discussion and Conclusions}
\label{sec:disc}

The evidence for hot gas in low mass protostars that could give rise to radio free--free emission has been strengthened by recent analyses of Herschel-PACS data (van Kempen et~al. 2010) which have revealed spectral features indicative of small-scale shocks created along the cavity walls of protostellar outflows. Of these, C--type shocks give rise to H$_2$0 emission, and J--type shocks within the lower density jet give rise to observed {\sc Oi} features. Combinations of these shocks can also give rise to observed {\sc OH} emission. However, the gas in such shocks is heated to temperatures of up to only 3000\,K (van~Kempen et~al. 2010), just sufficient to produce free--free emission. Typical temperatures for free--free emission are in the range $4000\leq T_{\rm{e}} \leq 20000$\,K, with a mean value of approximately 8000\,K (Dickinson et~al. 2003). Below 5000\,K emission through molecular lines and bands becomes significant. Indeed the theoretical lower limit that we show in Fig.~\ref{fig:fcorr} from Anglada (1996) for the amount of free-free emission from shock ionisation for a given outflow force, $F$, assumes $T_{\rm{e}}=10^4$\,K. This relationship can be re-expressed to include a temperature dependence as $L_{\rm{rad}}\propto F_{\rm{out}} T_{\rm{e}}^{-0.45}$, implying that for a lower temperature such as that found in van~Kempen et~al. (2010) a smaller amount of radio emission would be expected for a given outflow force. This would make it more difficult to explain the observed radio luminosities as arising from shock ionisation.

It is unclear why free--free emission produced by shock ionisation in such regions should be correlated strongly with envelope mass but less so with outflow force, as has been observed in this work. Since the stellar mass, $M_{\ast}$, determines the outflow velocity it could be supposed that emission from shock ionisation in a molecular outflow might be correlated with $M_{\ast}$. If, as demonstrated by Chabrier \& Hennebelle (2010), the stellar mass is a linear function of the core mass then this may lead to a correlation between envelope mass and radio luminosity from shock ionisation. However, the formation of shocks along the length of an outflow is dependent not only on the velocity of that outflow, but also on the local environment. To establish a clear correlation would require not only uniformity of the local environment but also of the number of shocks formed by an outflow of particular velocity. In addition we note that the measured outflow force for protostellar objects is not necessarily a current value but instead represents the averaged historical accretion history of the source and therefore may not necessarily correspond to the radio luminosity, which depends on the ``current'' protostellar state. The radio emission from protostars is also known to possess a degree of variability, as observed in the VeLLO source L1014 (Shirley et~al. 2007), which may contribute to the spread in the correlations presented here. The data here are observed over only a single epoch, and a degree of decorrelation due to variability will be common to all the measured correlations.

Since the free--free emission (radio luminosity) observed here demonstrates a correlation with envelope mass, which is furthermore consistent with that expected from free--free emission as a consequence of partially ionised stellar winds (both spherical and collimated), it is possible that the thermal and non-thermal radio flux density (see e.g. Rodr{\'i}guez et~al. 1999) detected from low mass protostars arises as a consequence not only of different emission mechanisms, but also of different underlying astrophysical processes. In order to distinguish these mechanisms, which occur at separated locations within a protostellar system, much higher resolution observations are required than can be provided here.

In addition to free--free and synchrotron emission, we must also consider the possibility that the correlation of radio luminosity with mass is a consequence of the cm-wave emission being due to a population of very cold dust. To check this hypothesis, the emission observed is fitted by a modified blackbody representing a possible very cold dust component. However, in order to accommodate the amount of emission seen at 1.8\,cm a dust temperature of $T_{\rm{d}}<5$\,K and a value of $\beta<<1$ are required. Such a low cold thermal equilibrium temperature for big dust grains is unrealistic, as is the spectral dust emissivity index. Current observations suggest $\beta$ to be in the range 1--2.5 (e.g. Boulanger et al. 1996; Paradis et al. 2009) and the Kramers--K{\"o}nig equations also suggest a value of 1 as a theoretical lower limit to $\beta$ (Emerson 1988). The existence of very cold dust that would explain the observed centimetre emission therefore seems very unlikely.

In summary we find five main conclusions from this work:
\begin{itemize}
\item We find that 72\% of Class~0 objects in the extended sample have detected radio counterparts at 16\,GHz, compared with a lower detection rate of 27\% for Class~I objects. No starless cores were detected. This detection rate for Class~I sources differs from that of Paper~I and we hypothesise that the discrepancy may be due to environmental effects causing localised differences in the radio emission from Class I objects, whereas radio emission from Class 0 objects arises as a consequence of an intrinsic emission mechanism. 
\item These new data strengthen the correlations from Paper I of radio luminosity with bolometric and infra-red luminosity in the low luminosity limit and significantly increase the available radio data for low and very low luminosity protostars.
\item We find no improvement in the correlation of radio luminosity with outflow force from these new data. We suggest that for the low temperatures found in shocks along molecular outflow cavities the measured outflow forces cannot explain the observed radio luminosities.
\item We find a correlation of radio luminosity with core mass for Class 0 objects which does not hold for Class I objects. We suggest that this difference is broadly consistent with theories of episodic mass accretion, but requires additional data for Class~I sources in order to be confirmed.
\item We hypothesize that the observed correlation with core mass seen for Class 0 objects may be due to the predicted relationship between stellar mass loss rate and radio luminosity, as suggested by Panagia \& Felli (1975) and Reynolds (1986). With the currently available data it is not possible to distinguish the case of an ionised spherical stellar wind from a collimated outflow, however this correlation suggests that such a mechanism is more likely to account for the observed free--free radio emission than the alternatively hypothesised mechanism of shock ionisation.
  
\end{itemize}

\section{ACKNOWLEDGEMENTS}
We thank the staff of the Lord's Bridge Observatory for their
invaluable assistance in the commissioning and operation of the
Arcminute Microkelvin Imager. We also thank the anonymous referee for their careful reading and useful comments on this paper. The AMI-LA is supported by Cambridge
University and the STFC. CRG, TS, TF, MO and MS   
acknowledge the support of PPARC/STFC studentships. YP acknowledges support from the Cambridge Commonwealth Trust and the Cavendish Laboratory. AS would like to acknowledge support from Science 
Foundation Ireland under grant 07/RFP/PHYF790.

\appendix

\section{Source List}

Sources detected in the AMI-LA fields. Column [1] indicates the AMI-LA source designation in increasing Right Ascension, column [2] indicates the field in which the source was detected, column [3] lists the Right Ascension in J2000 coordinates, column [4] lists the source Declination in J2000 coordinates, column [5] lists the peak flux density of the source at 16\,GHz in units of $\mu$Jy\,beam$^{-1}$ and column [6] lists the integrated flux density in units of mJy; column [7] lists known associations for each source. Sources are designated as ``radio'' where they have a counterpart in the NVSS catalogue (Condon et~al. 1998).
\begin{table*}
\caption{ Detected sources in the Perseus region at 16\,GHz with the AMI-LA. \label{tab:moresources1}}
\begin{tabular}{ccccccc}
\hline\hline
AMI & field & RA & dec & $S_{\rm{p}}$ & $S_{\rm{int}}$ & assoc. \\
 & & (J2000) & (J2000) & ($\mu$Jy\,beam$^{-1}$) & (mJy) & \\
\hline
01& DCE08-055 & 03 25 22.3 & +30 45 10.8 & 470 & $670\pm39$ & [H07]-30 \\
02& DCE08-055 & 03 25 31.6 & +30 44 35.8 & 174 & $396\pm27$ & \\
03& DCE08-055 & 03 25 36.2 & +30 45 20.8 & 1713 & $2059\pm105$ & DCE08-055 \\
04& DCE08-055 & 03 25 38.9 & +30 44 00.8 & 470 & $537\pm33$ & DCE08-056 \\
\emph{03}& DCE08-056 & 03 25 36.4 & +30 45 18.1 & 1822 & $2045\pm105$ & DCE08-055 \\
\emph{04}& DCE08-056 & 03 25 38.7 & +30 44 03.1 & 510 & $519\pm34$ & DCE08-056 \\
05& DCE08-063 & 03 27 29.8 & +30 14 48.8 & 2649 & $3178\pm161$ & radio \\
06& DCE08-063 & 03 27 41.7 & +30 11 43.8 & 3265 & $4337\pm218$ & radio \\
07& DCE08-065 & 03 28 38.3 & +31 06 06.8 & 154 & $288\pm25$ & DCE08-065 \\
08& DCE08-071 & 03 28 56.6 & +31 14 20.9 & 273 & $320\pm25$ & [H07]-44 \\
09& DCE08-071 & 03 28 57.4 & +31 14 12.1 & 658 & $1211\pm63$ &  \\
10& DCE08-071 & 03 29 00.2 & +31 12 05.7 & 239 & $239\pm22$ & DCE08-071 \\
11& DCE08-071 & 03 29 10.7 & +31 13 30.7 & 1685 & $1689\pm87$ & [H07]-41 \\
12& DCE08-071 & 03 29 11.8 & +31 13 16.0 & 439 & $486\pm31$ & DCE08-073 \\
13& DCE08-073 & 03 29 03.1 & +31 13 41.6 & 240 & $240\pm22$ & VLA15? \\
\emph{11}& DCE08-073 & 03 29 10.5 & +31 13 31.6 & 1657 & $1733\pm90$ & [HO7]-41 \\
14& DCE08-073 & 03 29 12.1 & +31 13 01.6 & 468 & $516\pm37$ & DCE08-073 \\
15& DCE08-081 & 03 30 43.5 & +30 27 21.5 & 561 & $590\pm36$ & \\
16& DCE08-084 & 03 31 19.0 & +30 47 25.2 & 9276 & $9580\pm480$ & radio\\
17& DCE08-084 & 03 31 21.0 & +30 45 15.2 & 276 & $324\pm34$ & DCE08-084 \\
18& DCE08-084 & 03 31 24.9 & +30 44 15.2 & 285 & $445\pm37$ & \\
19& DCE08-088 & 03 32 17.6 & +30 49 47.6 & 346 & $424\pm38$ & DCE08-088 \\
20& DCE08-088 & 03 32 23.4 & +30 49 42.6 & 648 & $712\pm48$ & radio \\
21& DCE08-090 & 03 32 25.3 & +31 05 10.9 & 5324 & $5408\pm272$ & radio \\
22& DCE08-090 & 03 32 29.2 & +31 02 35.9 & 102 & $117\pm26$ & DCE08-090 \\
23& DCE08-092 & 03 33 13.7 & +31 07 07.6 & 207 & $207\pm29$ & DCE08-092 \\
24& DCE08-092 & 03 33 16.1 & +31 06 57.6 & 215 & $235\pm29$ & DCE08-093 \\
25& DCE08-092 & 03 33 16.8 & +31 08 02.6 & 138 & $138\pm28$ & [H07]-7 \\
26& DCE08-092 & 03 33 18.0 & +31 09 32.6 & 449 & $490\pm36$ & [H07]-1 \\  
27& DCE08-092 & 03 33 20.7 & +31 09 12.6 & 258 & $283\pm30$ & \\
28& DCE08-092 & 03 33 21.1 & +31 07 32.6 & 515 & $684\pm44$ & [H07]-2 \\
29& DCE08-092 & 03 33 27.3 & +31 07 07.6 & 442 & $629\pm41$ & [H07]-10 \\
\emph{23}& DCE08-093 & 03 33 13.7 & +31 07 07.6 & 207 & $207\pm24$ & DCE08-092 \\
\emph{24}& DCE08-093 & 03 33 16.1 & +31 06 57.6 & 243 & $243\pm25$ & DCE08-093 \\
\emph{25}& DCE08-093 & 03 33 16.8 & +31 08 02.6 & 138 & $138\pm23$ & [H07]-7 \\
\emph{26}& DCE08-093 & 03 33 18.0 & +31 09 32.6 & 419 & $424\pm31$ & [H07]-1 \\  
\emph{27}& DCE08-093 & 03 33 20.7 & +31 09 12.6 & 267 & $304\pm27$ & \\
\emph{28}& DCE08-093 & 03 33 21.1 & +31 07 32.6 & 548 & $640\pm39$ & [H07]-2 \\
\emph{29}& DCE08-093 & 03 33 27.3 & +31 07 07.6 & 424 & $582\pm36$ & [H07]-10 \\
30& DCE08-105 & 03 43 49.8 & +32 00 32.9 & 215 & $343\pm29$ &  \\
31& DCE08-105 & 03 43 56.2 & +32 00 52.9 & 207 & $535\pm35$ & DCE08-105 \\
\emph{30}& DCE08-106 & 03 43 51.7 & +32 00 54.7 & 143 & $201\pm22$ & \\
\emph{31}& DCE08-106 & 03 43 55.7 & +32 00 50.3 & 218 & $472\pm31$ & DCE08-105 \\
32& DCE08-106 & 03 43 58.0 & +32 03 09.7 & 137 & $162\pm22$ & DCE08-106 \\
33& DCE08-060 & 03 26 28.6 & +30 16 18.1 & 6321& $7153\pm358$ & radio \\
34& DCE08-080 & 03 29 52.2 & +31 39 01.1 & 84$^{\ast}$ & $90\pm19$ & DCE08-080 \\
35& DCE08-104 & 03 43 43.5 & +32 04 52.9 & 324 & $1543\pm79$ & RIDGE \\
36& DCE08-104 & 03 43 53.0 & +32 02 12.9 & 134 & $135\pm20$ & \\
37& DCE08-104 & 03 43 57.3 & +32 05 27.9 & 485 & $529\pm33$ & \\
38& DCE08-104 & 03 43 56.5 & +32 00 52.9 & 395 & $646\pm38$ & [H07]-12 \\
39& DCE08-104 & 03 43 57.3 & +32 03 07.9 & 326 & $552\pm34$ & [H07]-13 \\
40& DCE08-104 & 03 43 59.7 & +32 03 07.9 & 249 & $275\pm24$ & IC348BN \\
\emph{38}& DCE08-107 & 03 43 56.5 & +32 00 49.9 & 433 & $982\pm54$ & [H07]-12 \\
\emph{39}& DCE08-107 & 03 43 57.7 & +32 03 09.9 & 360 & $534\pm35$ & [H07]-13 \\
41& DCE08-107 & 03 44 07.1 & +32 03 59.9 & 258 & $1146\pm61$ & \\
42& DCE08-107 & 03 44 13.0 & +32 01 24.9 & 229 & $229\pm25$ & [H07]-101 \\
43& DCE08-107 & 03 44 13.8 & +32 00 59.9 & 341 & $341\pm28$ & \\
44& DCE08-109 & 03 44 20.5 & +32 01 59.2 & 2010 & $2022\pm104$ &  \\
\hline
\end{tabular}
\begin{minipage}{\textwidth}
$^{\ast}$ Detected at 4.4\,$\sigma$.\\
\end{minipage}
\end{table*}
\begin{table*}
\caption{ Detected sources outside the Perseus region at 16\,GHz with the AMI-LA. \label{tab:moresources2}}
\begin{tabular}{ccccccc}
\hline\hline
AMI & field & RA & dec & $S_{\rm{p}}$ & $S_{\rm{int}}$ & assoc. \\
 & & (J2000) & (J2000) & ($\mu$Jy\,beam$^{-1}$) & (mJy) & \\
\hline
45& DCE08-005 & 04 41 12.7 & +25 46 35.4 & 523 & $645\pm41$ & DCE08-005 \\
46& DCE08-044 & 22 30 33.2 & +75 14 18.9 & 116 & $167\pm25$ & DCE08-044 \\
47& DCE08-045 & 22 30 48.6 & +75 12 47.2 & 127 & $159\pm17$ & radio \\
48& DCE08-045 & 22 31 05.6 & +75 13 37.2 & 125 & $179\pm17$ & DCE08-045 \\
49& DCE08-045 & 22 31 09.5 & +75 15 07.2 & 127 & $146\pm17$ & \\
50& DCE08-045 & 22 31 33.1 & +75 13 57.1 & 184 & $189\pm18$ & \\
51& DCE08-049 & 22 38 46.4 & +75 11 33.0 & 854 & $1173\pm64$ & DCE08-048/049 \\
\hline
\end{tabular}
\end{table*}

\section{AMI-LA Maps}
\label{app:maps}
Maps of each field are shown. AMI-LA data is shown uncorrected for the primary beam response as greyscale and contours of 5, 6, 7, 8\,$\sigma$ etc to 1\,mJy\,beam$^{-1}$ and then in increments of 1\,mJy\,beam$^{-1}$. The AMI-LA primary beam response is shown as a solid circle and the PSF for each dataset as a filled ellipse in the bottom left corner. The positions of \emph{Spitzer} embedded protostellar cores (Dunham et~al. 2008) are shown as unfilled squares, and the positions of Class 0, Class I and starless cores from Hatchell et~al. (2007) are indicated as crosses (``+''), stars and unfilled circles, respectively.
\begin{figure*}
\centerline{\includegraphics[width=0.4\textwidth]{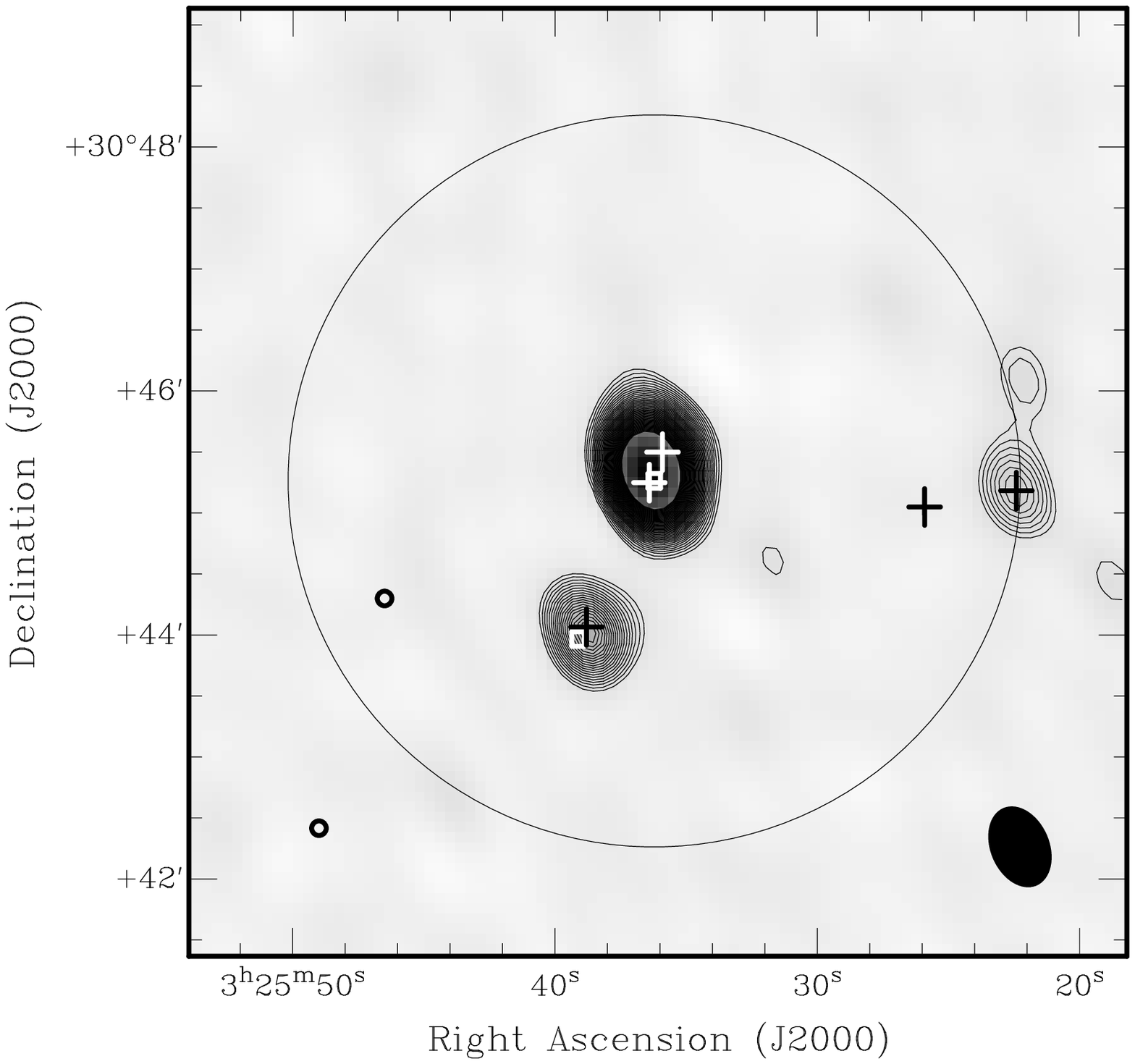}\qquad\includegraphics[width=0.4\textwidth]{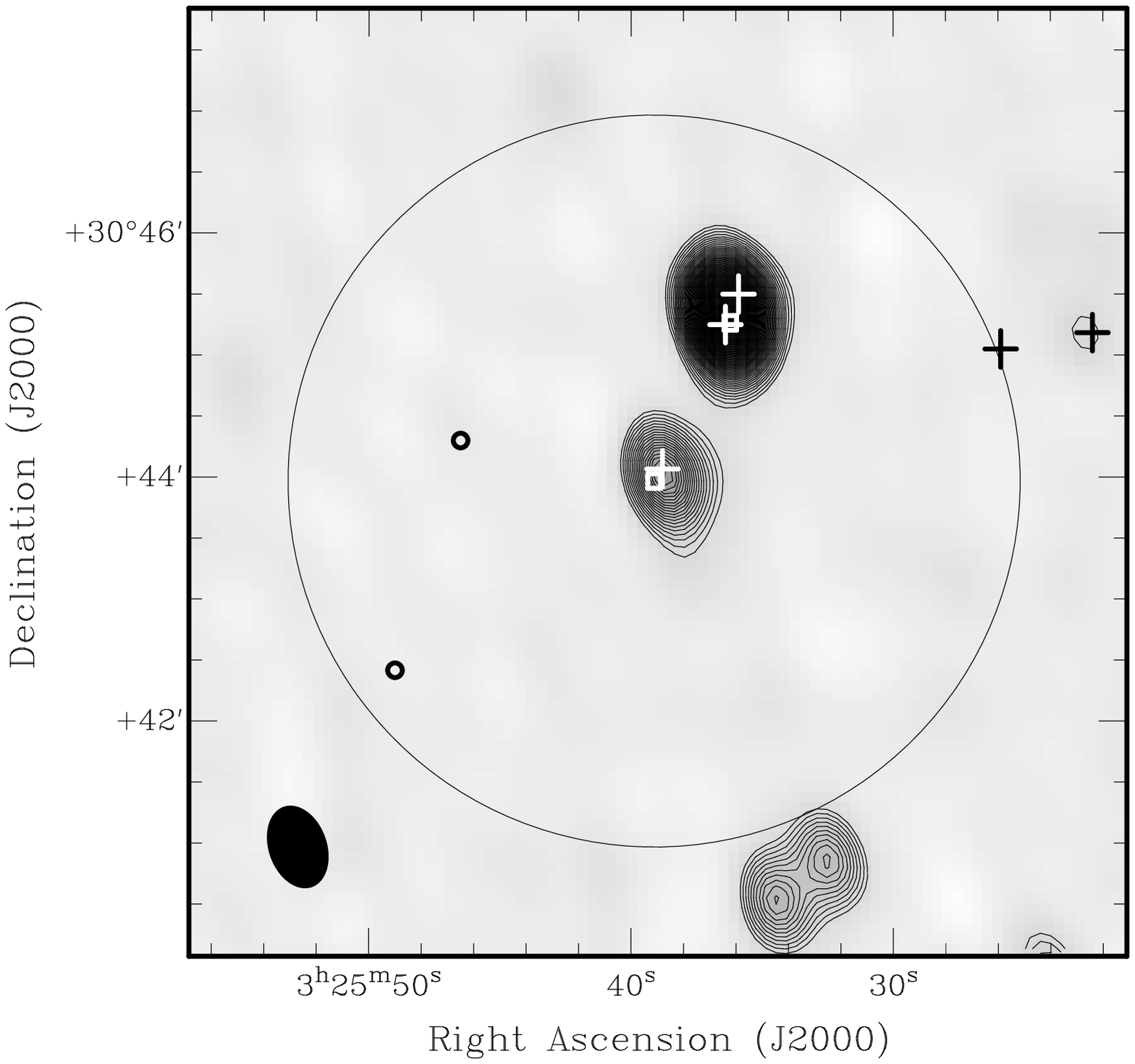}}

\centerline{DCE08-055\hspace{0.4\textwidth}DCE08-056}

\centerline{\includegraphics[width=0.4\textwidth]{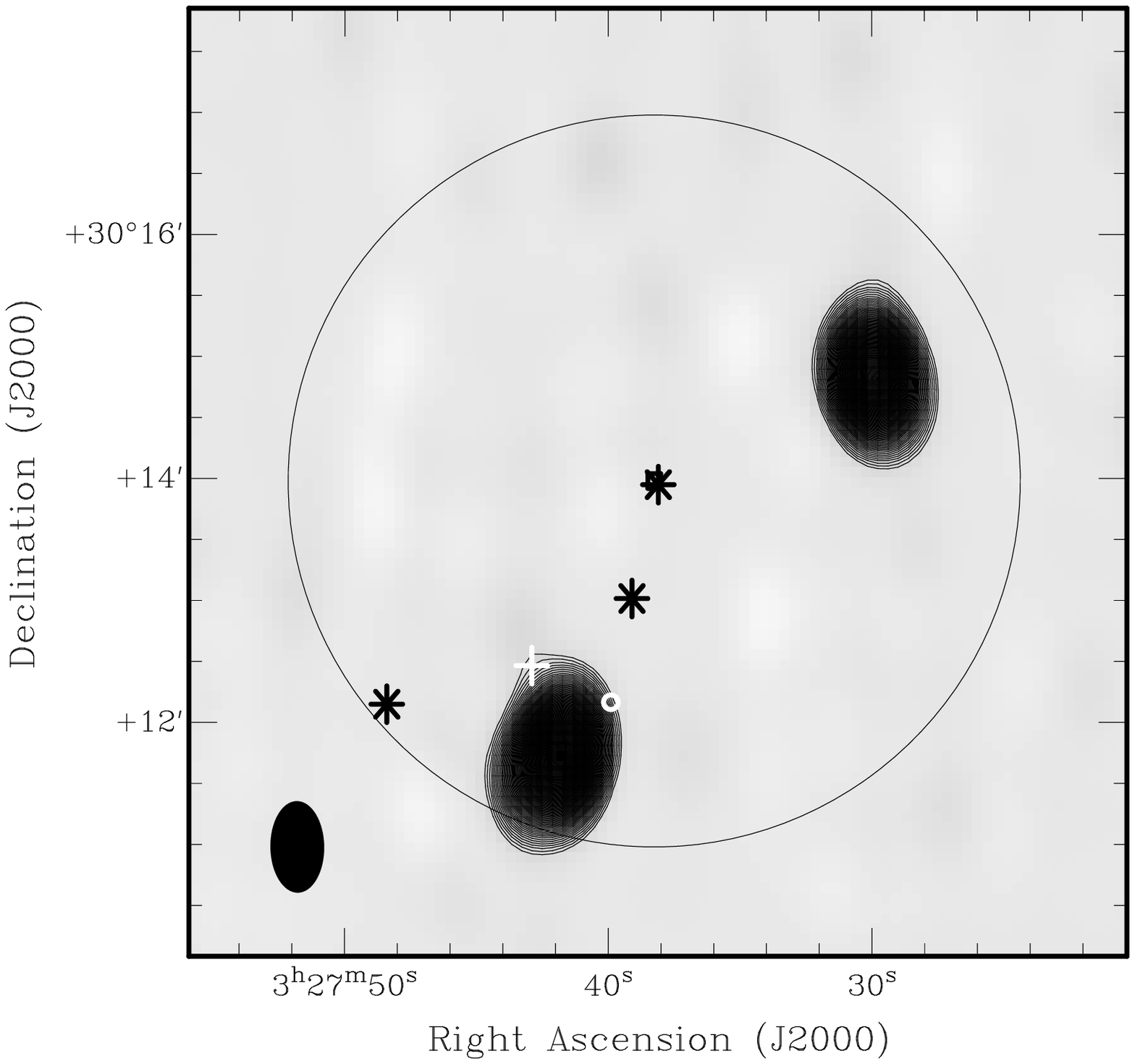}\qquad\includegraphics[width=0.4\textwidth]{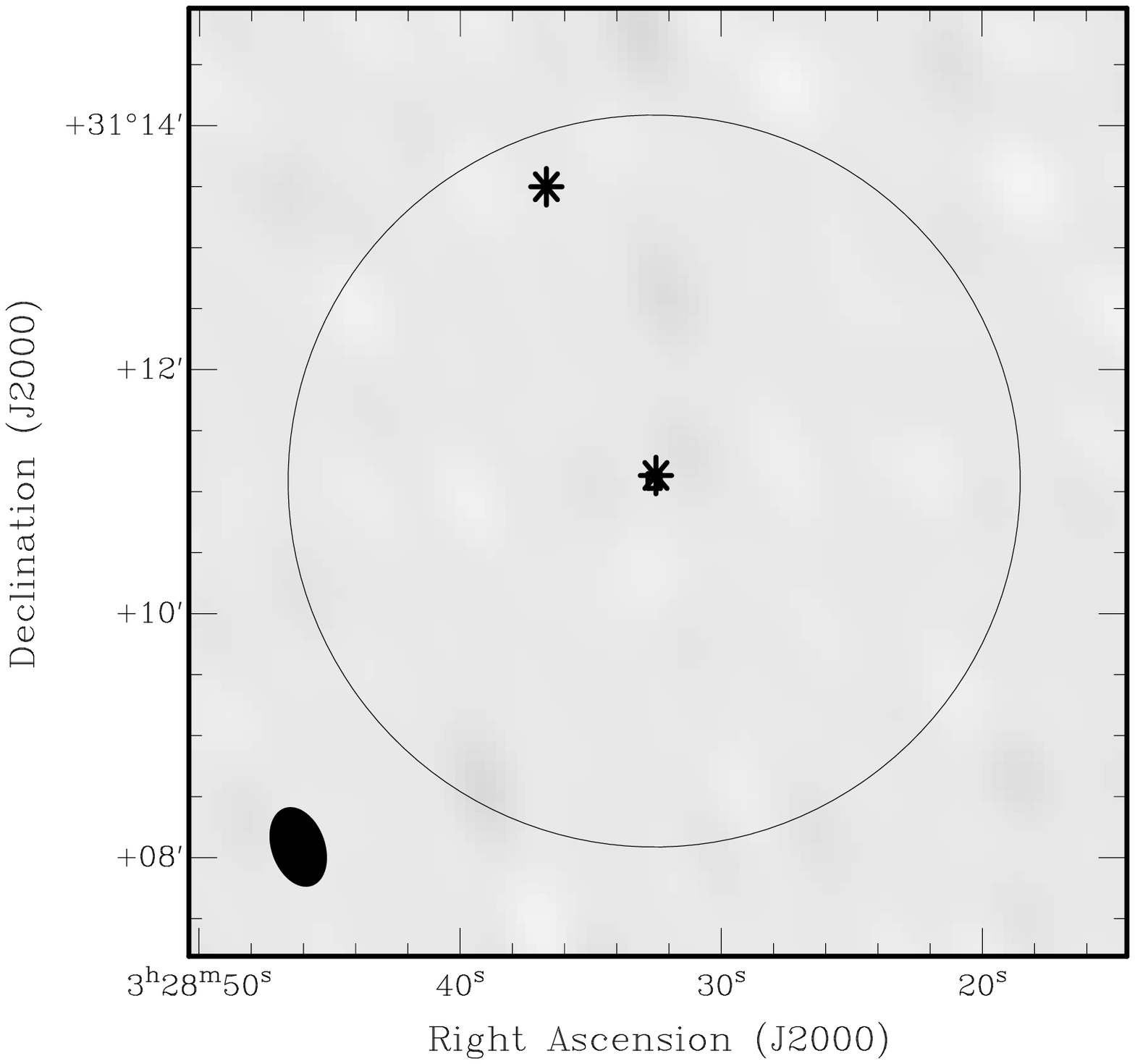}}

\centerline{DCE08-063\hspace{0.4\textwidth}DCE08-064}

\centerline{\includegraphics[width=0.4\textwidth]{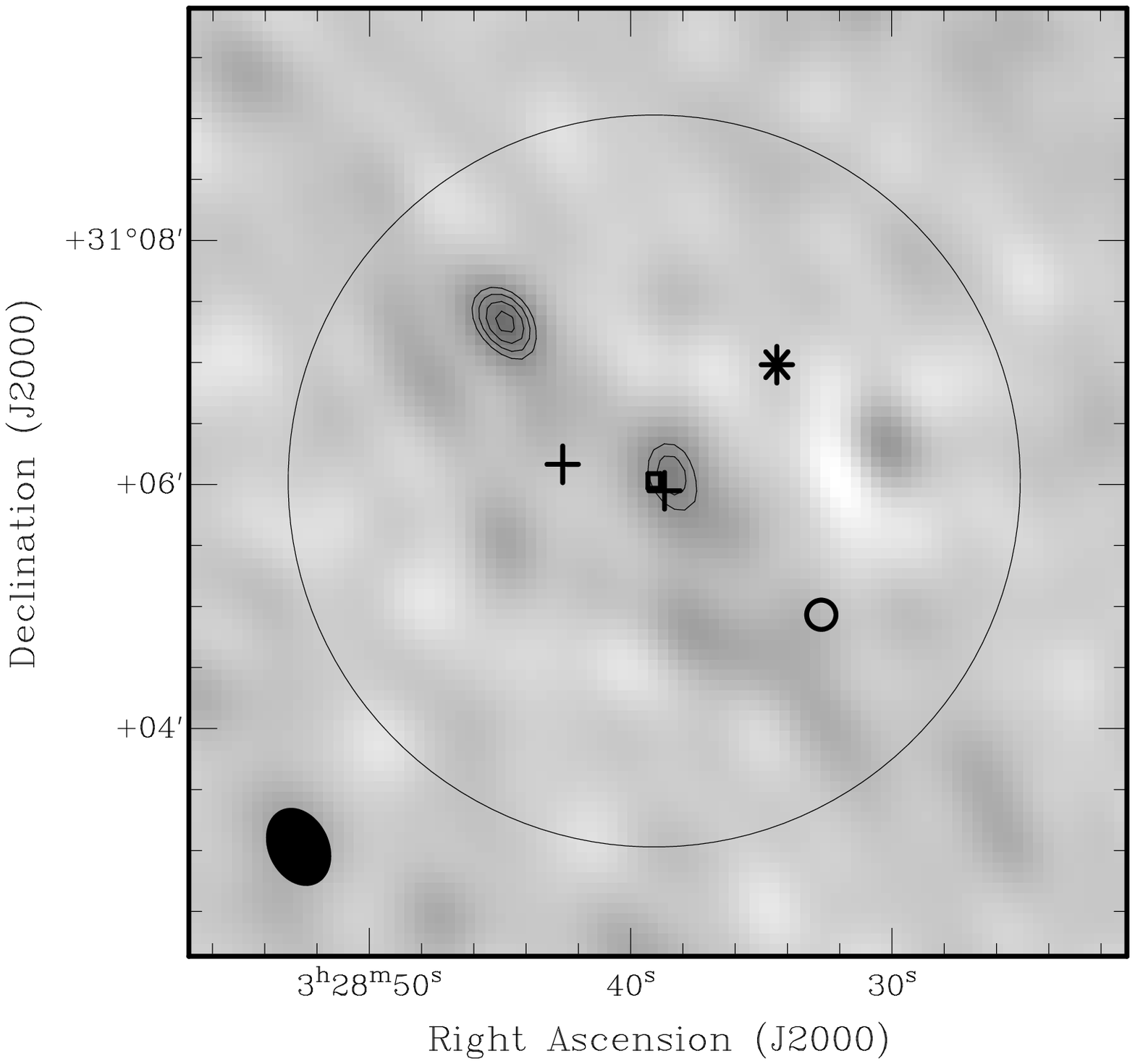}\qquad\includegraphics[width=0.4\textwidth]{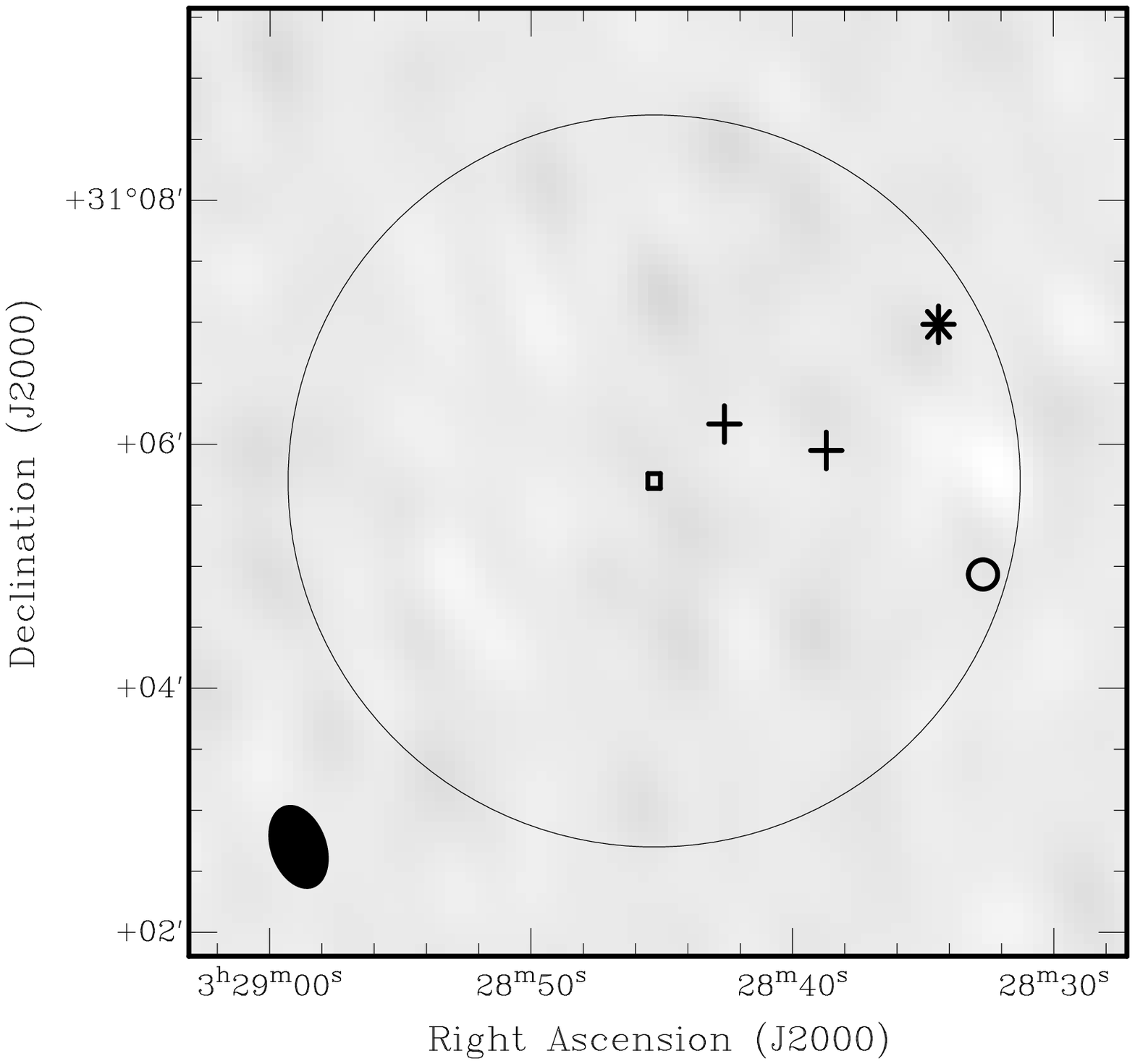}}

\centerline{DCE08-065\hspace{0.4\textwidth}DCE08-068}
\caption{\label{fig:maps}}
\end{figure*}
\begin{figure*}
\centerline{\includegraphics[width=0.4\textwidth]{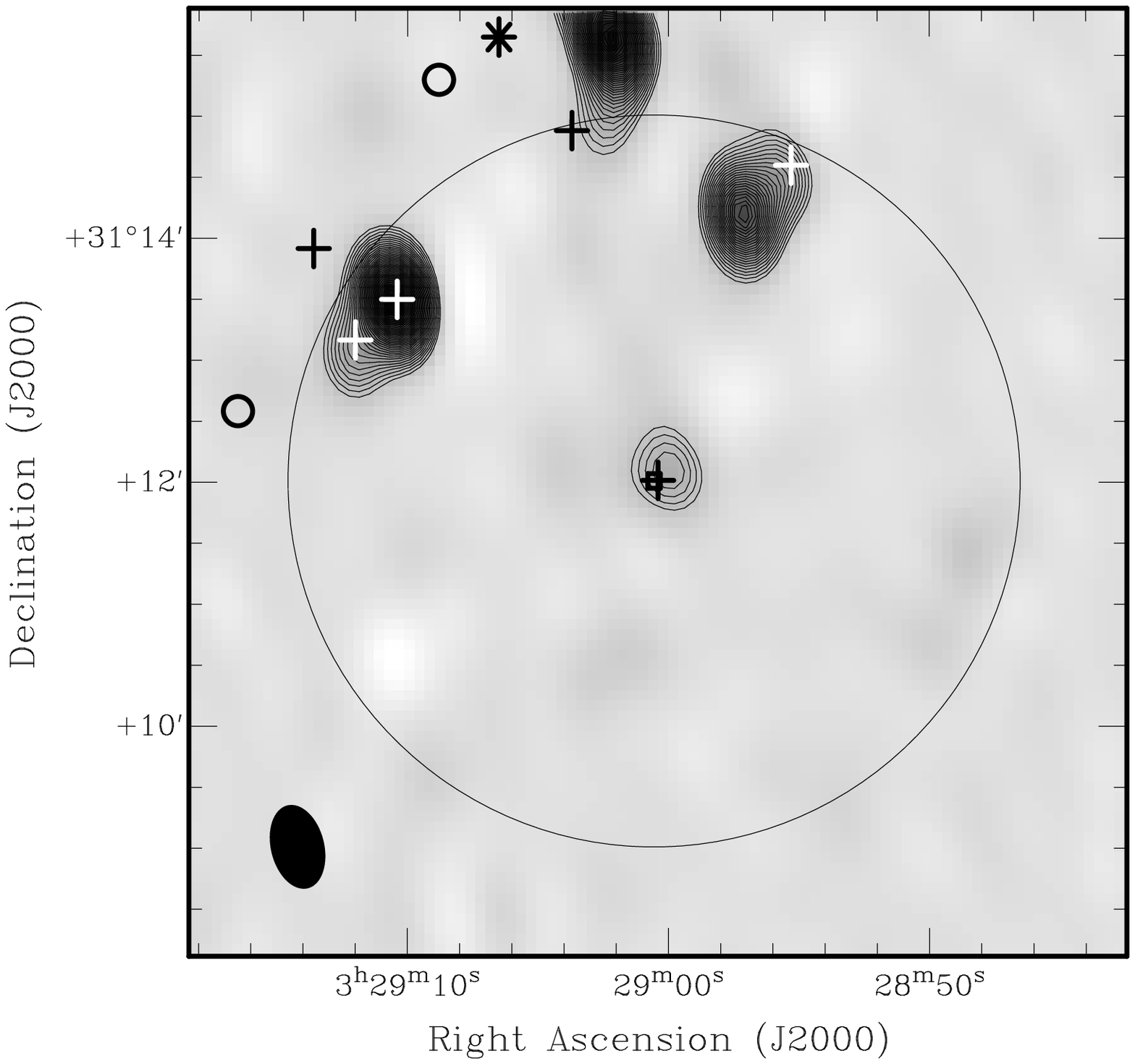}\qquad\includegraphics[width=0.4\textwidth]{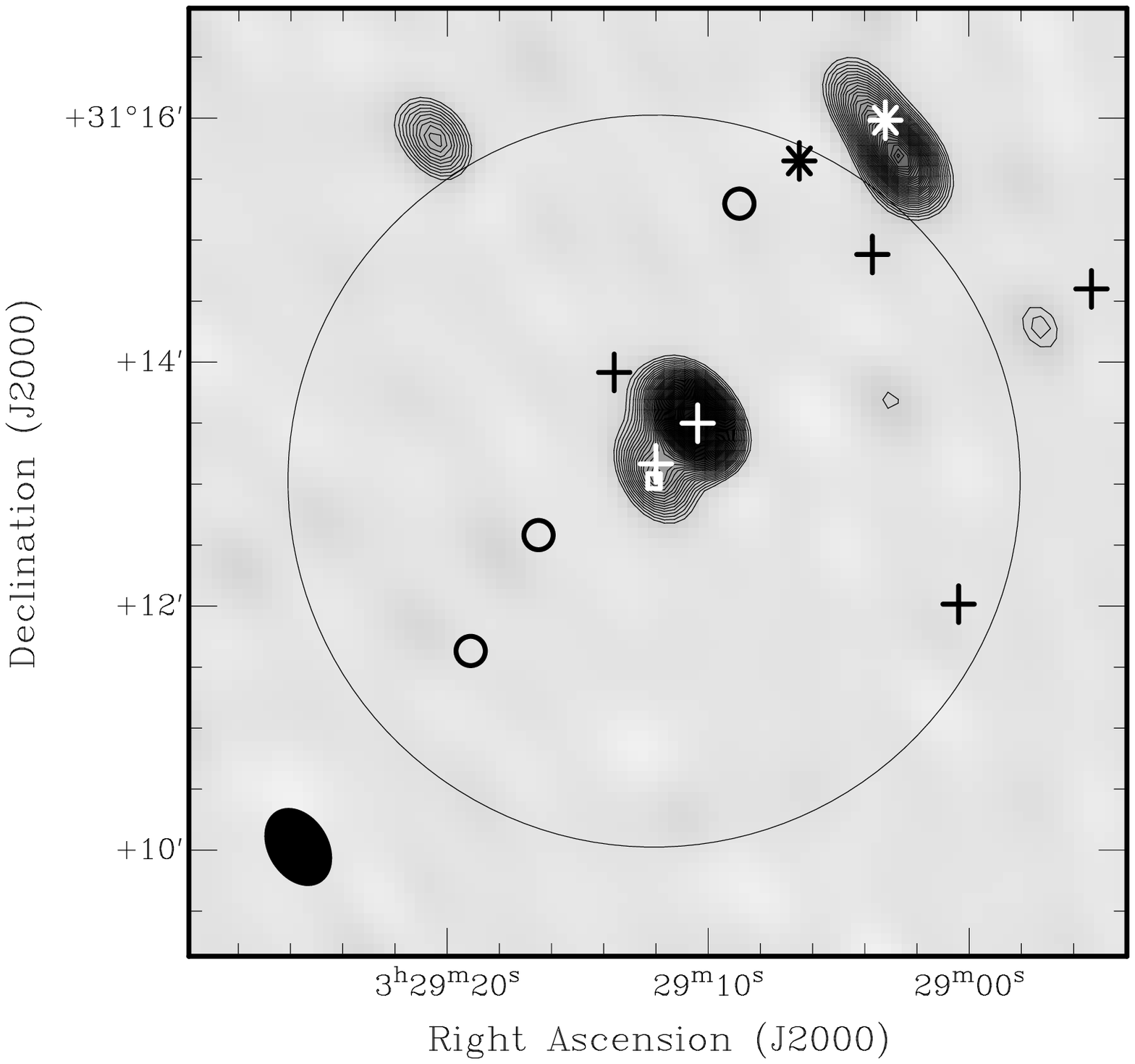}}

\centerline{DCE08-071\hspace{0.4\textwidth}DCE08-073}

\centerline{\includegraphics[width=0.4\textwidth]{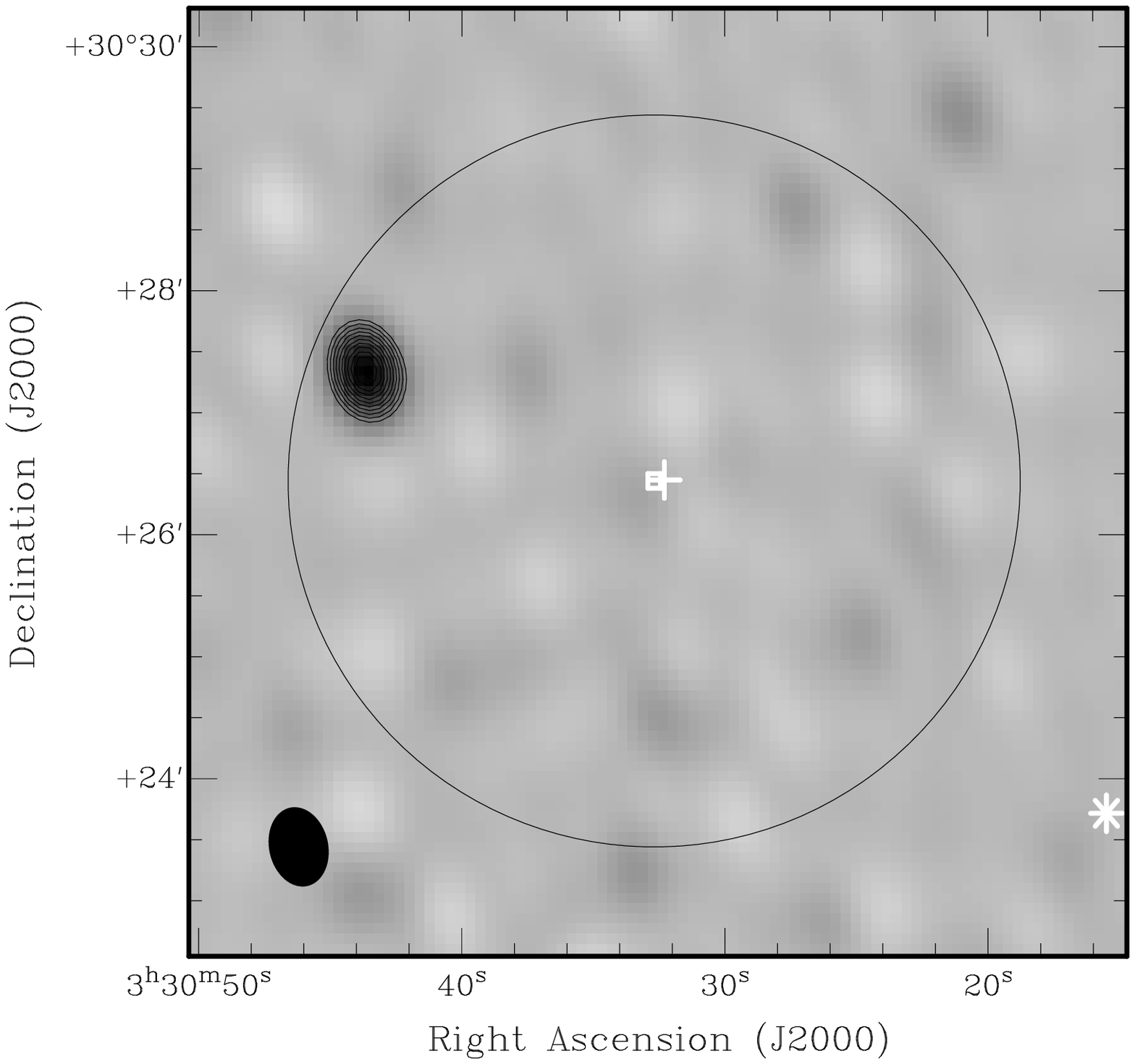}\qquad\includegraphics[width=0.4\textwidth]{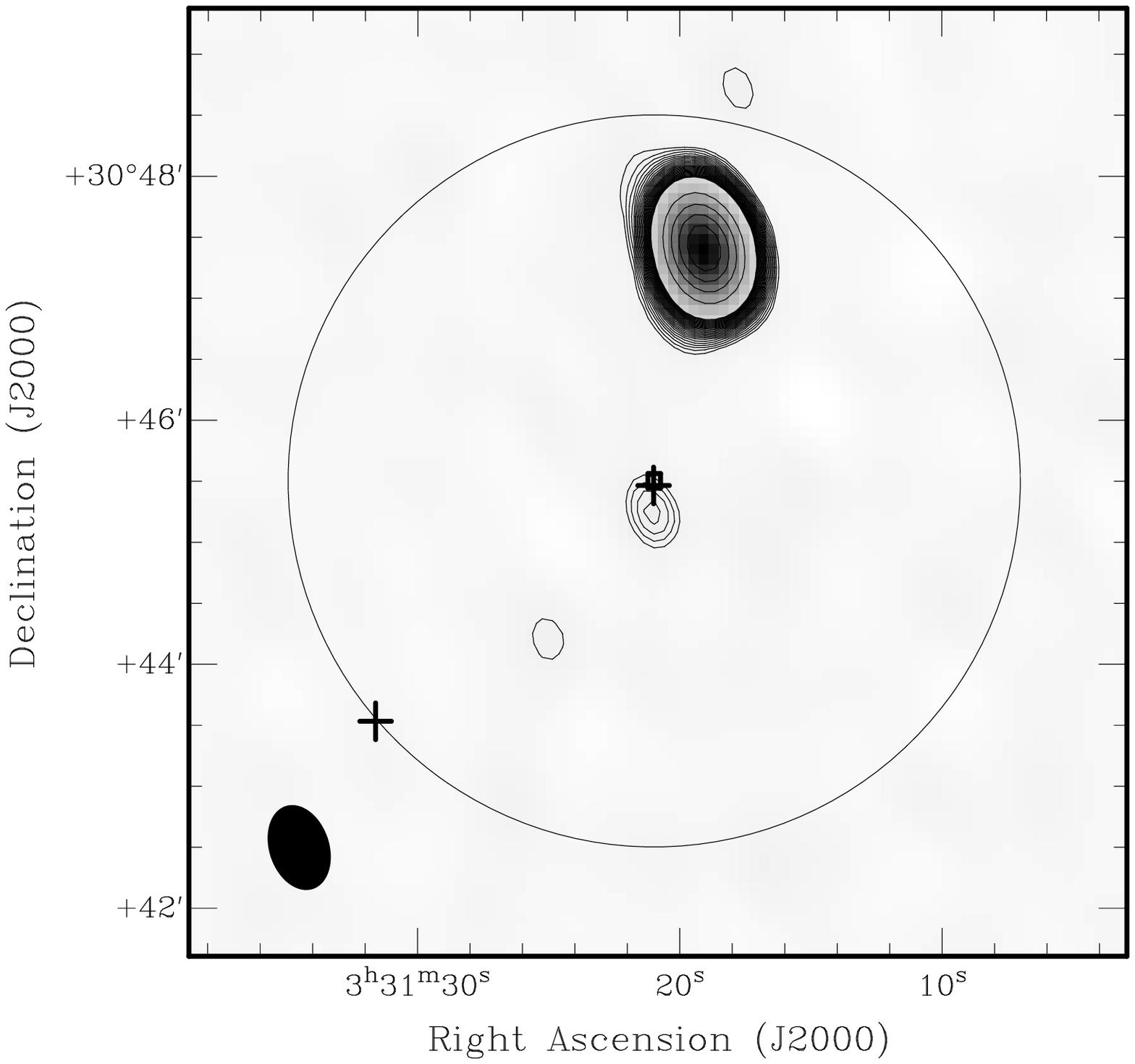}}

\centerline{DCE08-081\hspace{0.4\textwidth}DCE08-084}

\centerline{\includegraphics[width=0.4\textwidth]{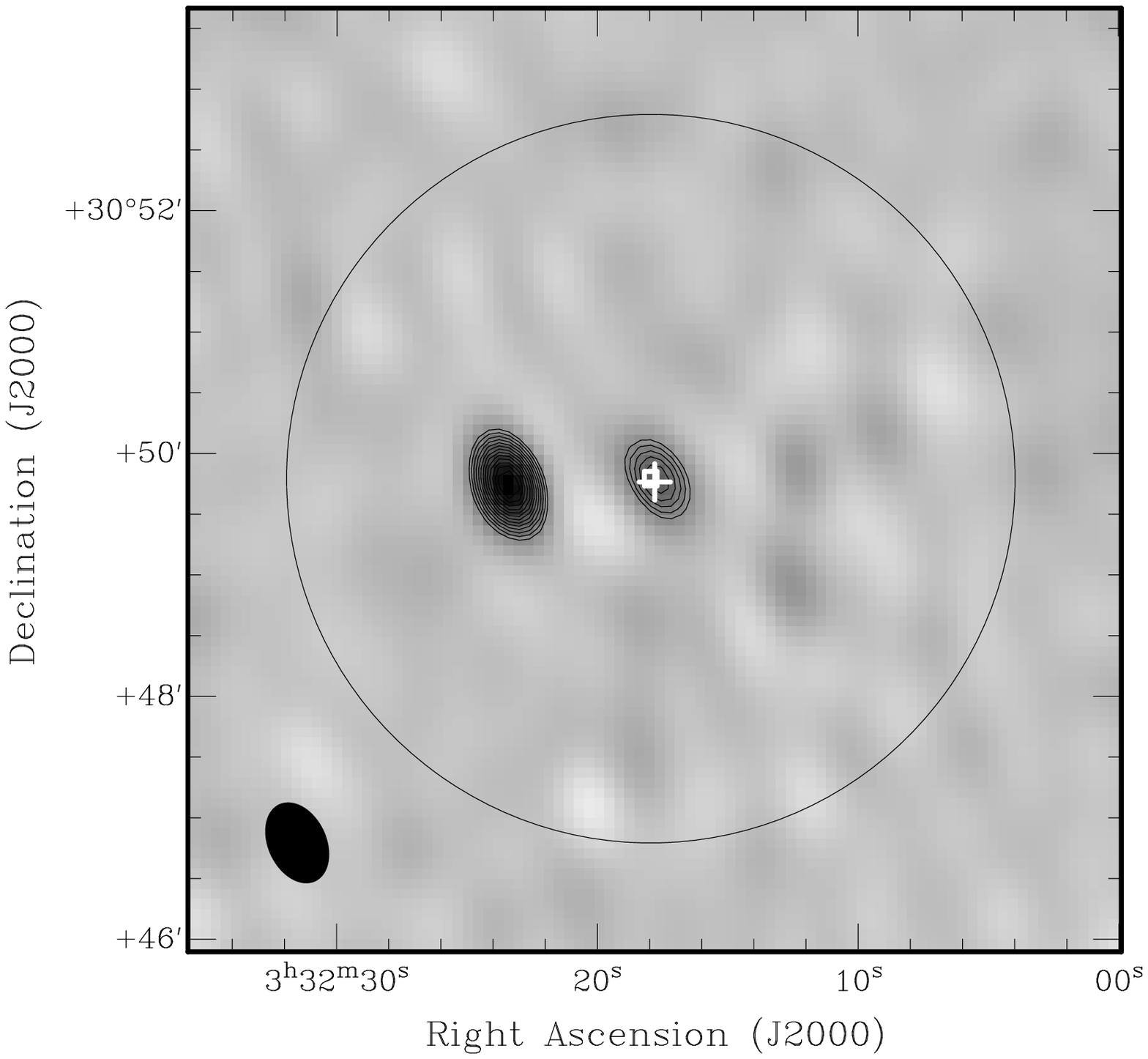}\qquad\includegraphics[width=0.4\textwidth]{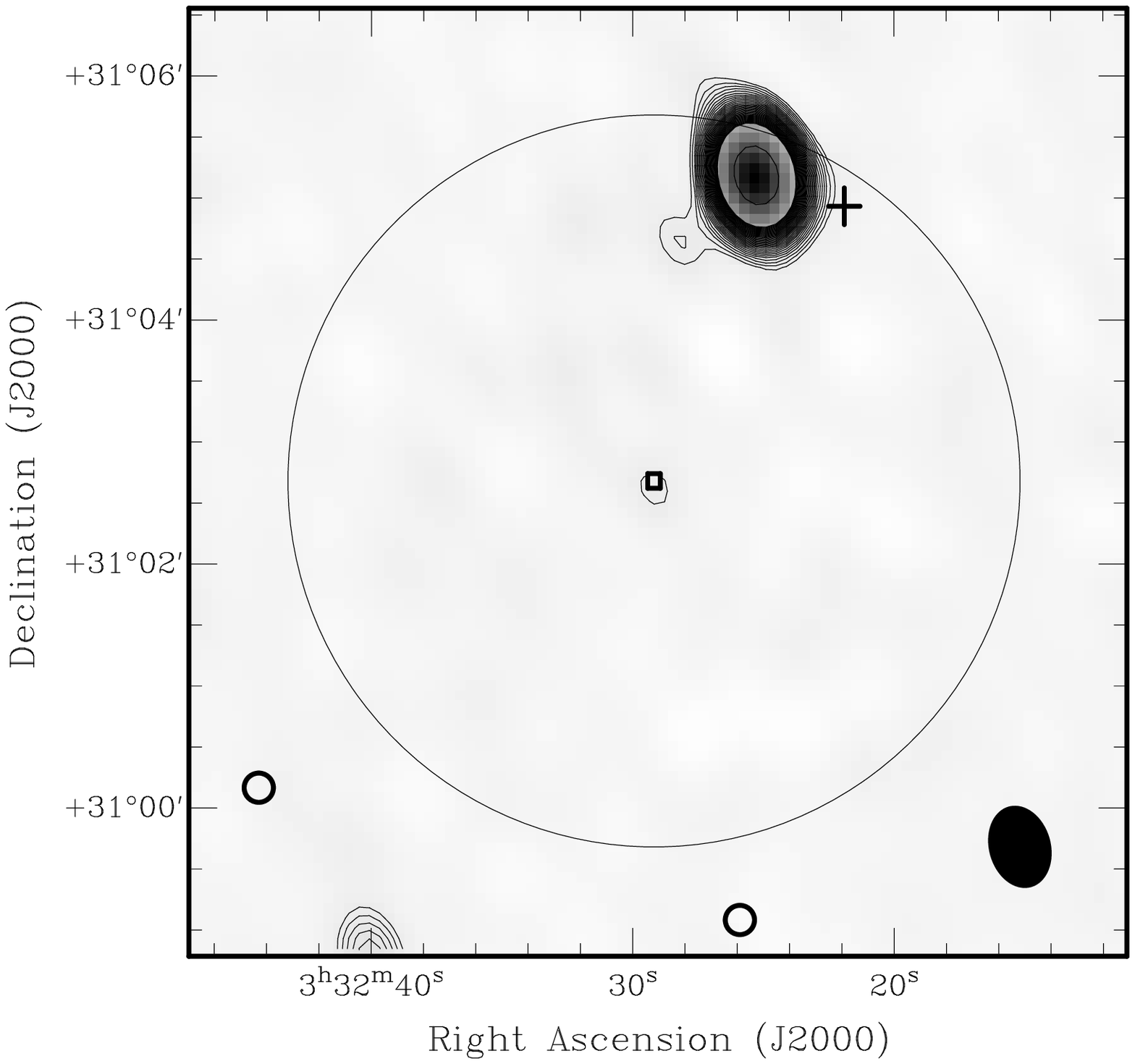}}

\centerline{DCE08-088\hspace{0.4\textwidth}DCE08-090}
\caption{\label{fig:maps}}
\end{figure*}
\begin{figure*}
\centerline{\includegraphics[width=0.4\textwidth]{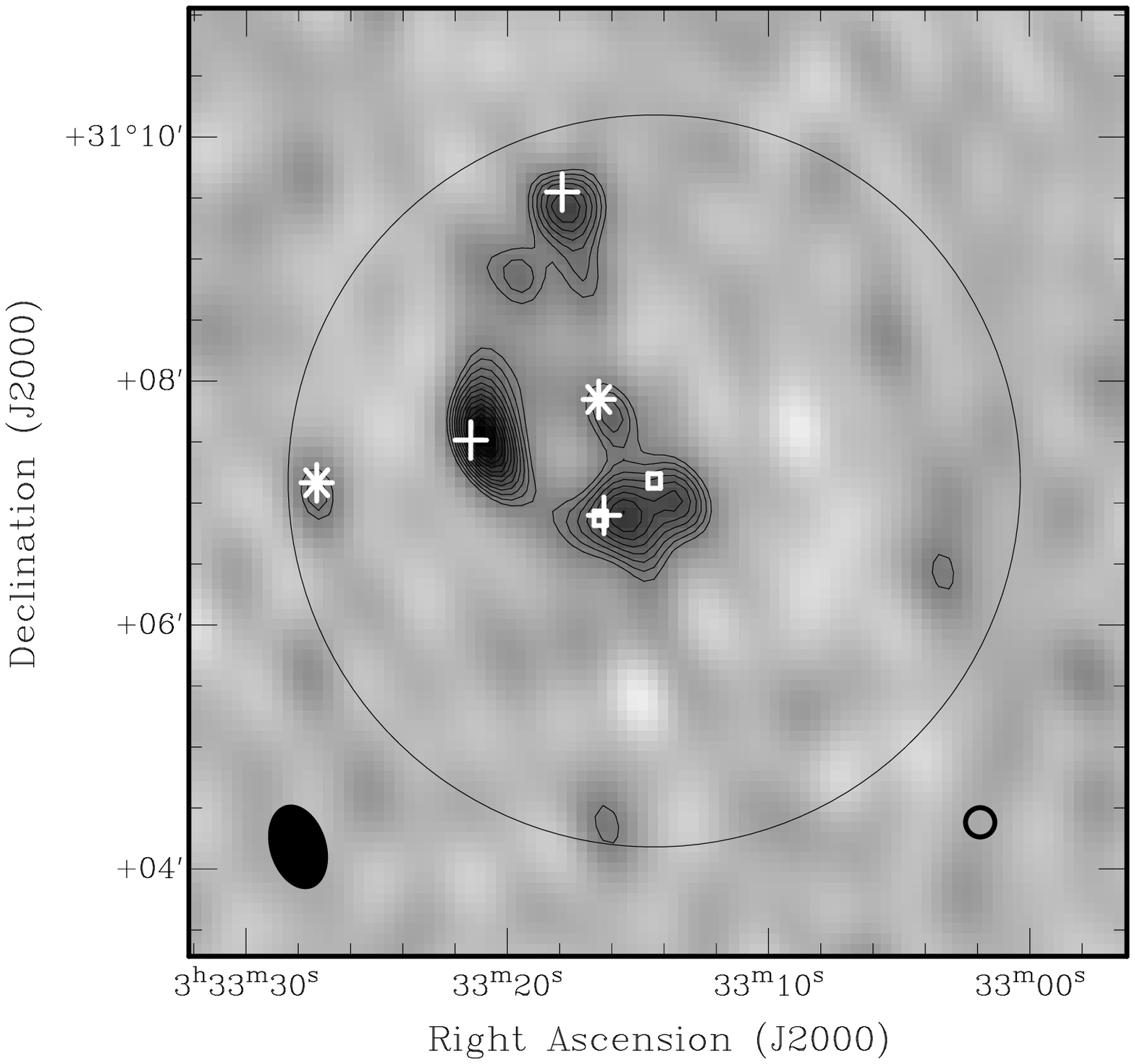}\qquad\includegraphics[width=0.4\textwidth]{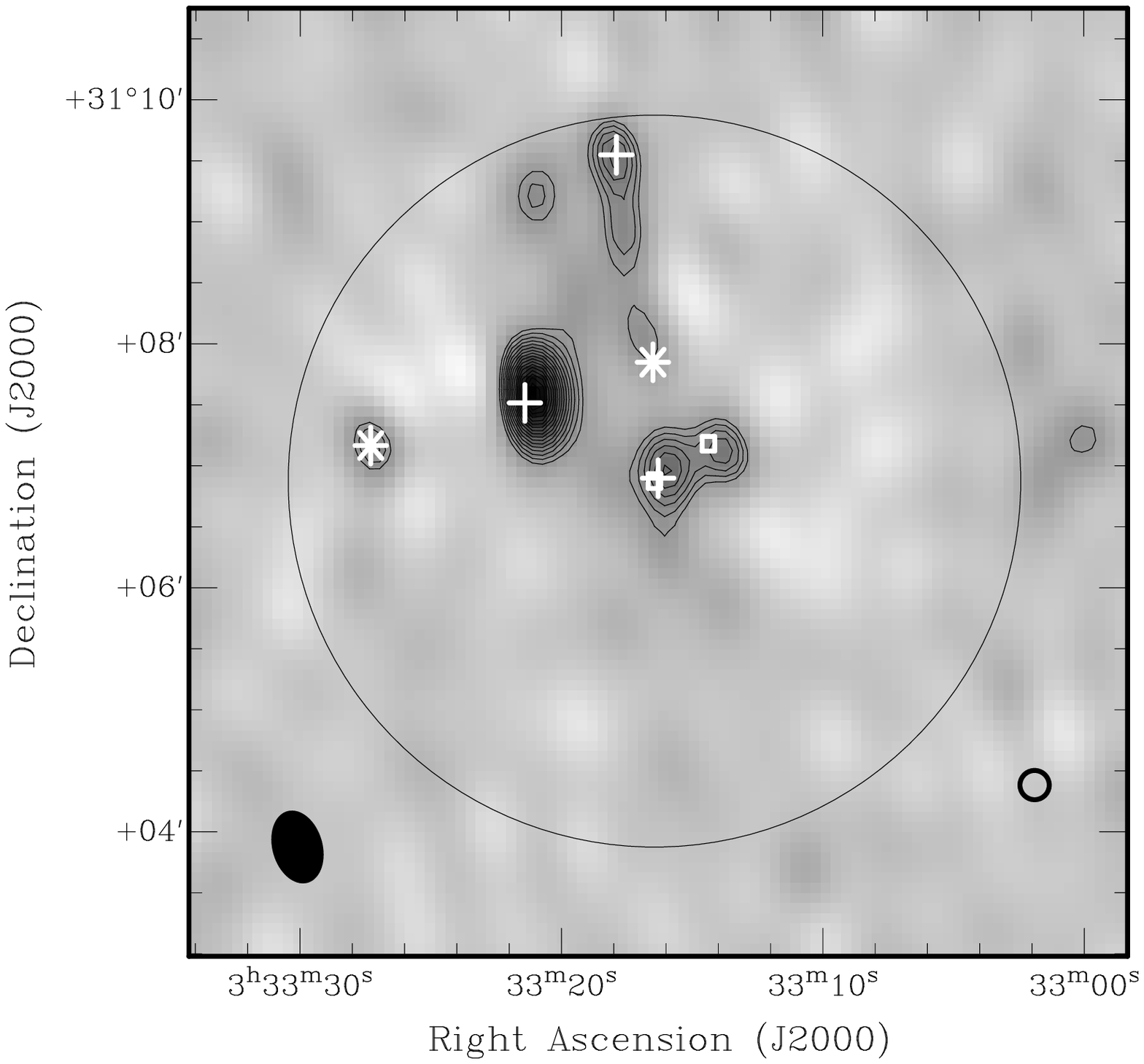}}

\centerline{DCE08-092\hspace{0.4\textwidth}DCE08-093}

\centerline{\includegraphics[width=0.4\textwidth]{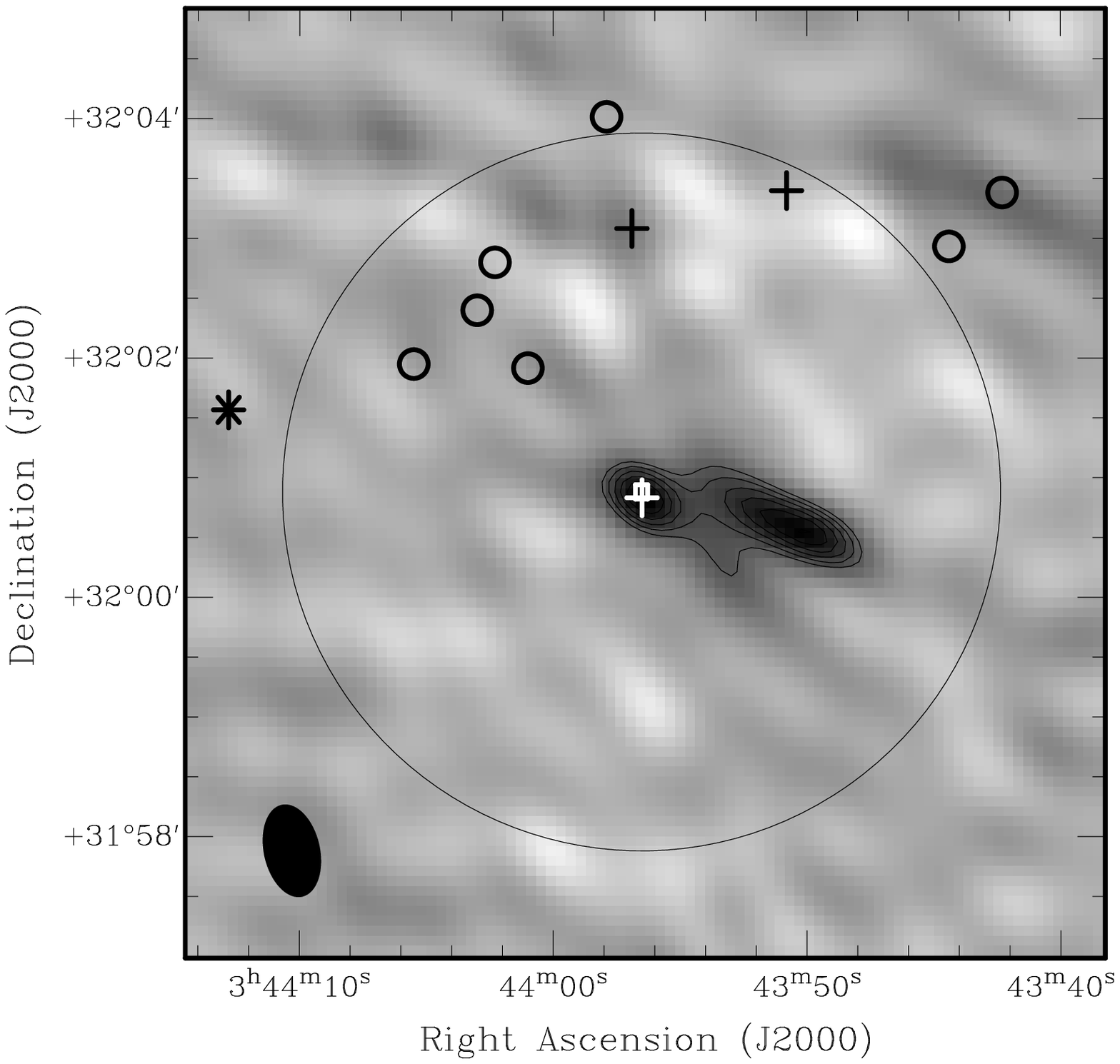}\qquad\includegraphics[width=0.4\textwidth]{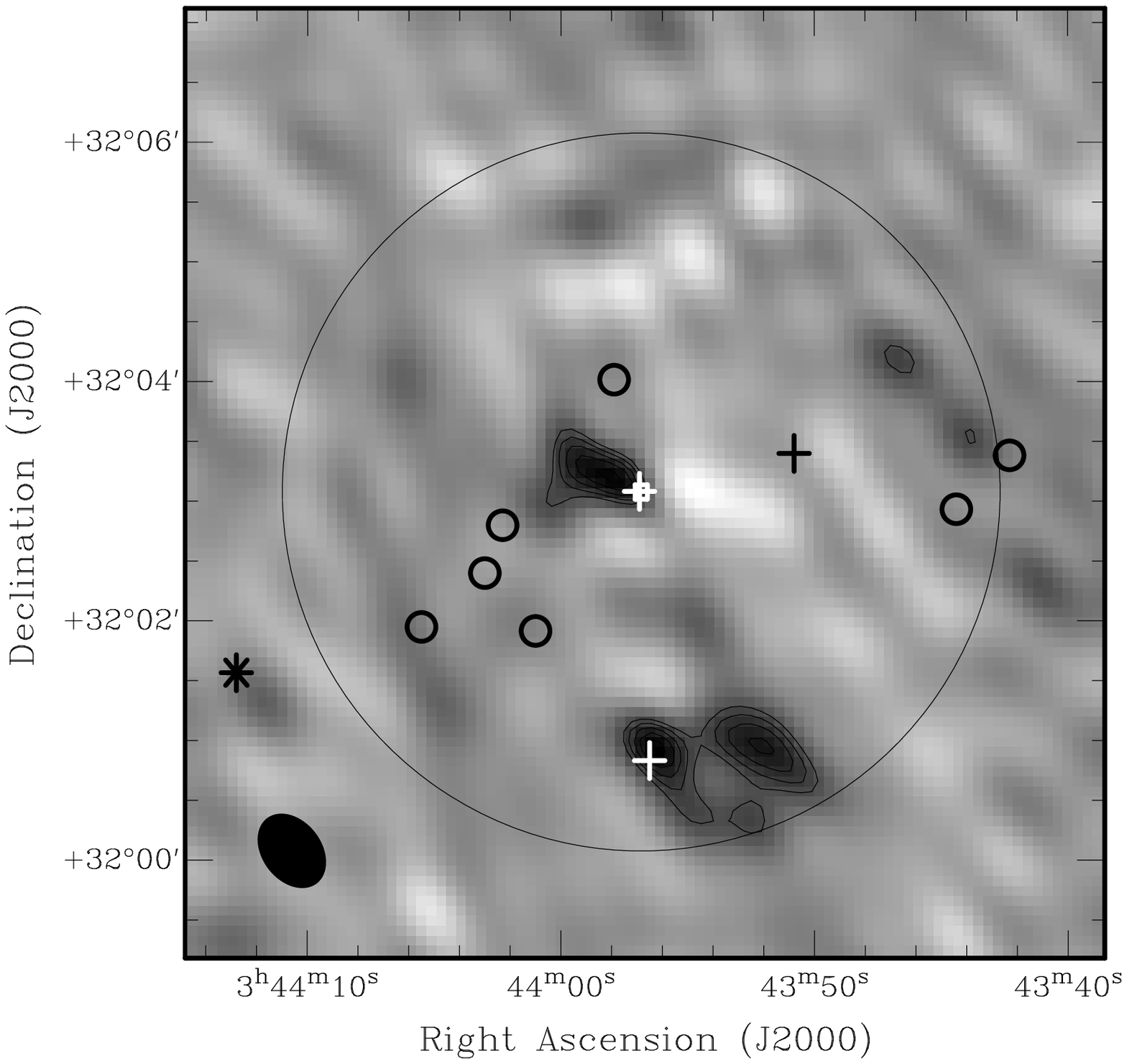}}

\centerline{DCE08-105\hspace{0.4\textwidth}DCE08-106}

\centerline{\includegraphics[width=0.4\textwidth]{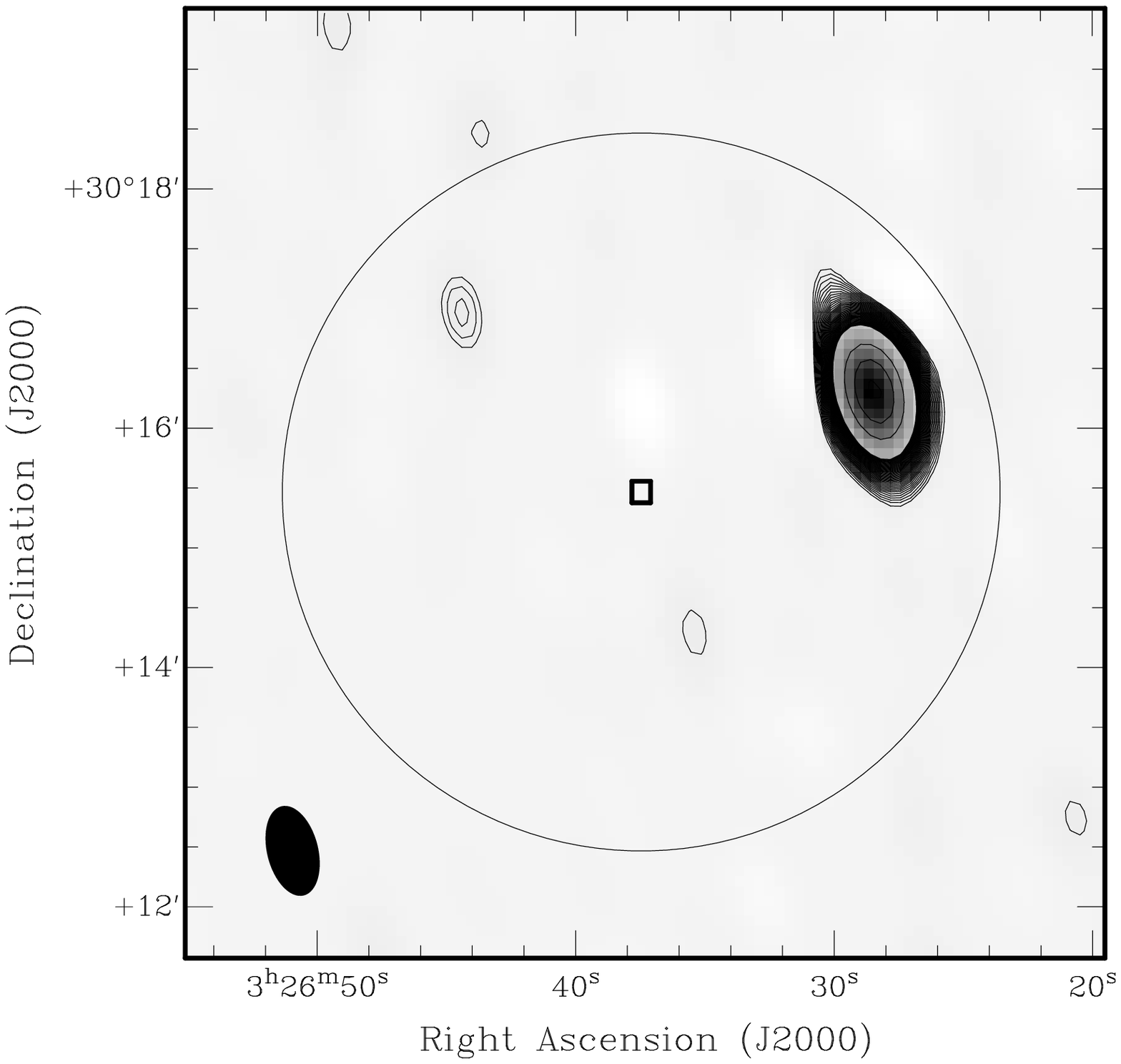}\qquad\includegraphics[width=0.4\textwidth]{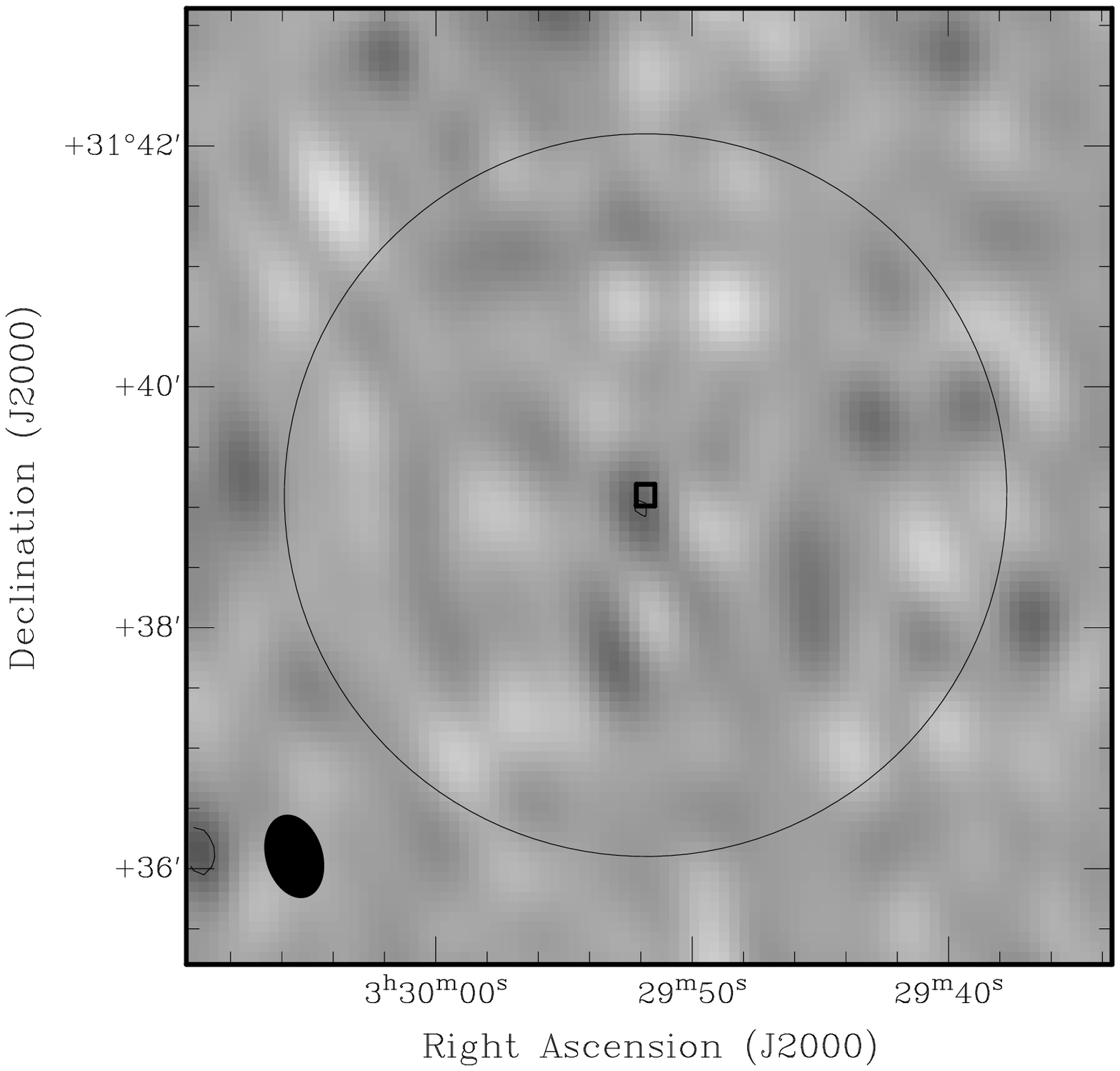}}

\centerline{DCE08-060\hspace{0.4\textwidth}DCE08-080}
\caption{\label{fig:maps}}
\end{figure*}
\begin{figure*}
\centerline{\includegraphics[width=0.4\textwidth]{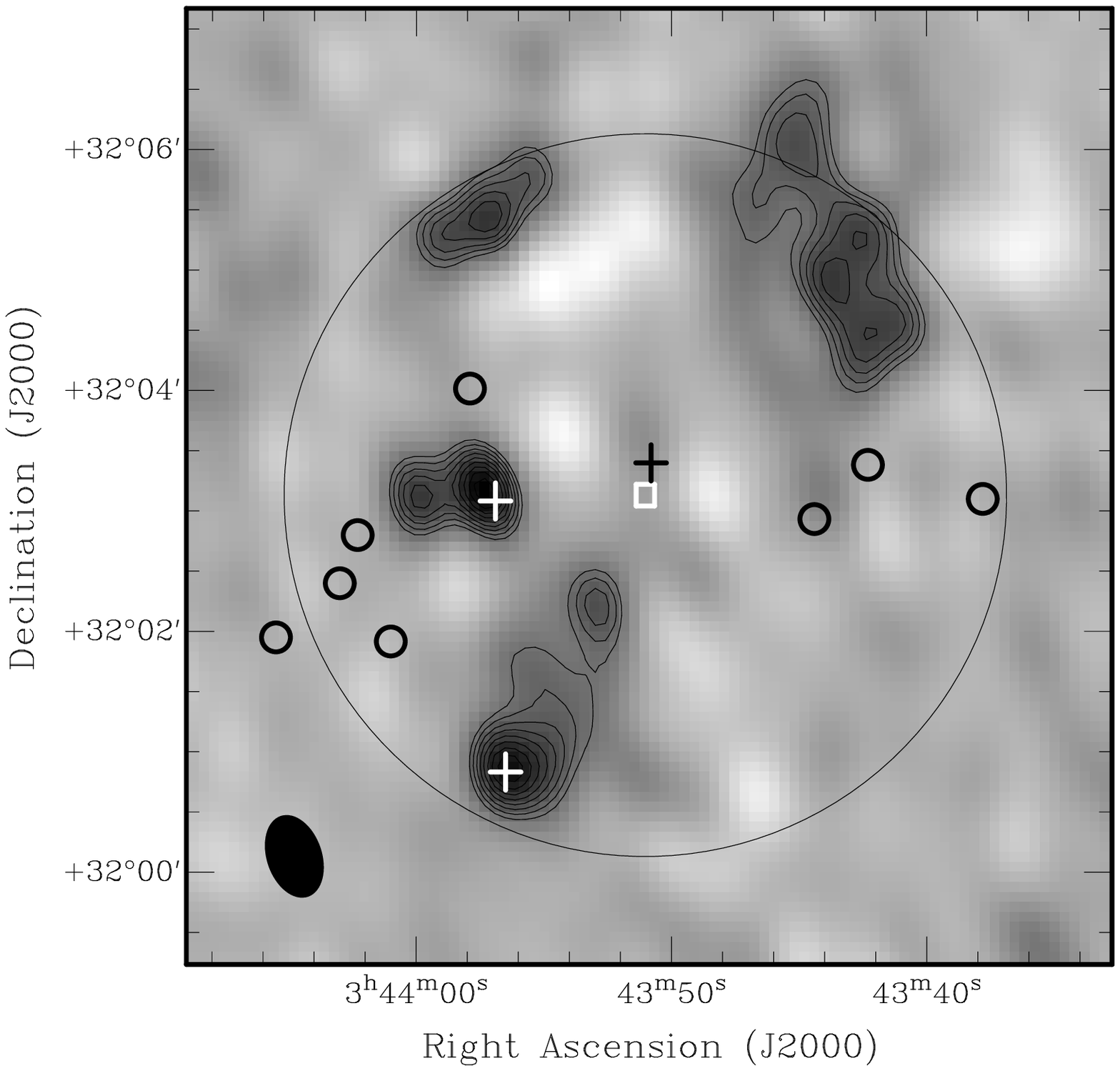}\qquad\includegraphics[width=0.4\textwidth]{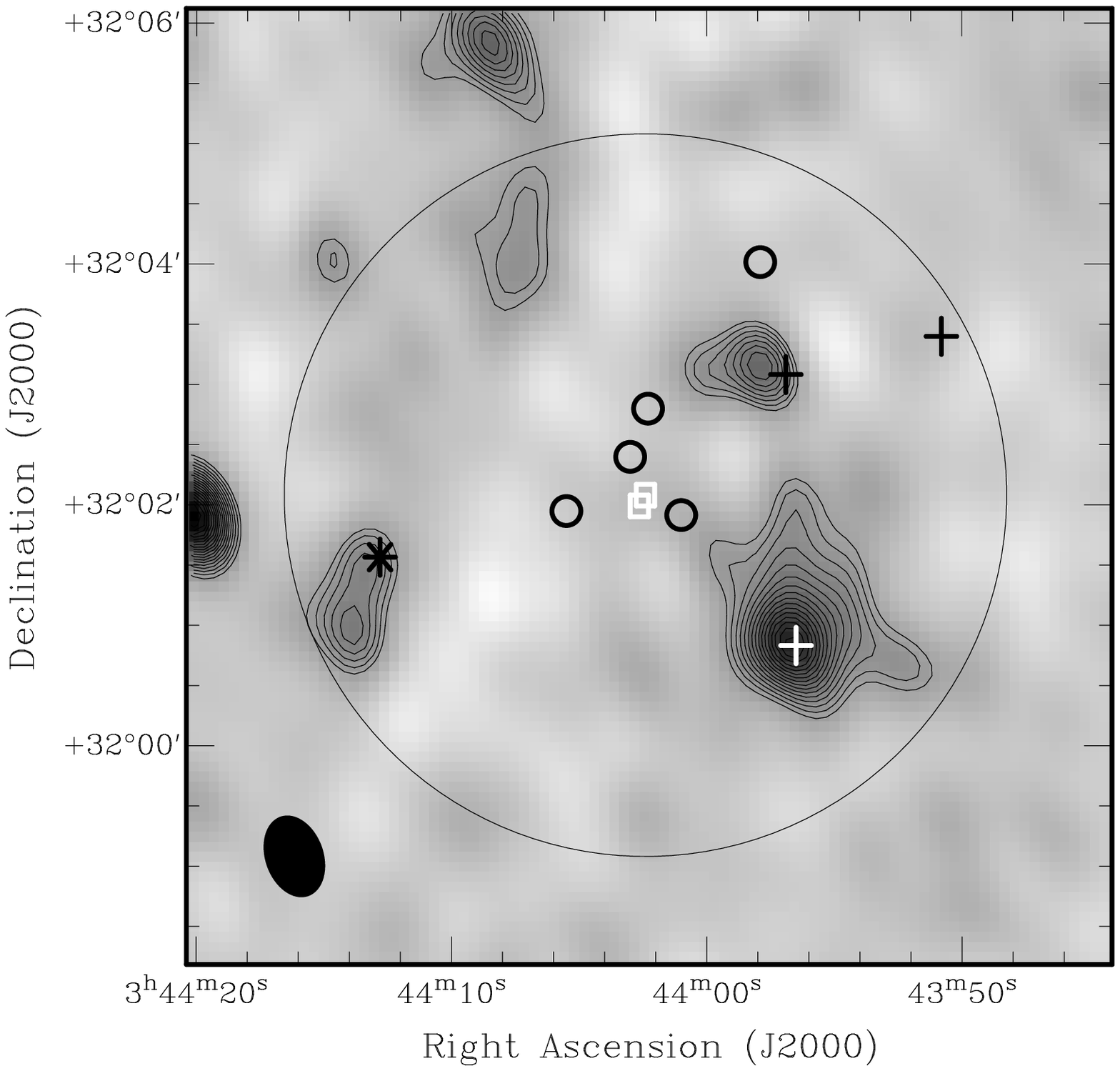}}

\centerline{DCE08-104\hspace{0.4\textwidth}DCE08-107}

\centerline{\includegraphics[width=0.4\textwidth]{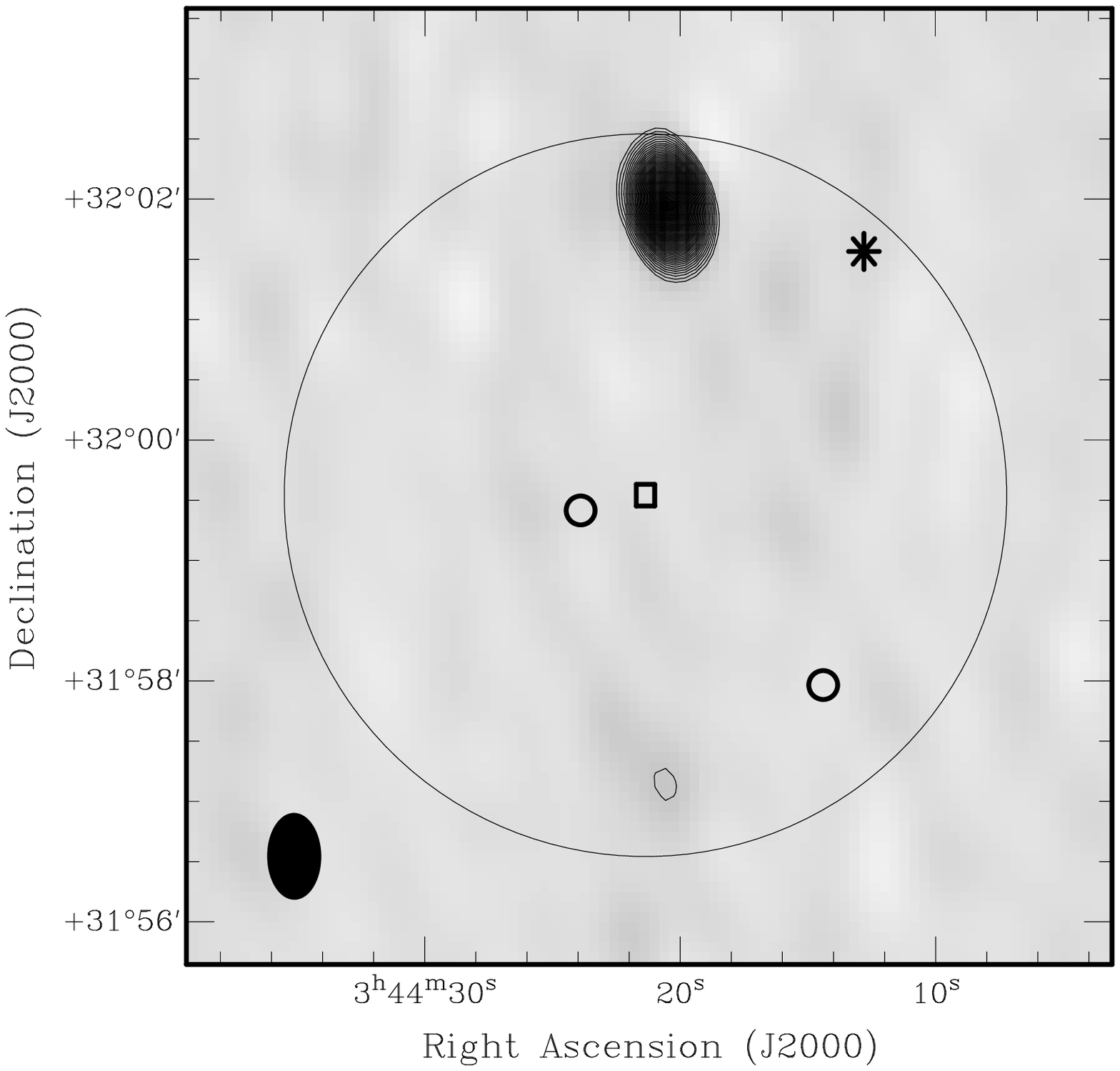}}

\centerline{DCE08-109}
\caption{\label{fig:maps}}
\end{figure*}


\begin{thebibliography}{}
\setlength{\labelwidth}{0pt}

\bibitem[\protect\citeauthoryear{Hurley-Walker et 
al.}{2009}]{2009MNRAS.398..249H} AMI Consortium: Hurley-Walker N., et al., 2009, MNRAS, 
398, 249 

\bibitem[\protect\citeauthoryear{AMI Consortium: Scaife et 
al.}{2011}]{2011MNRAS.410.2662A} AMI Consortium: Scaife, et al., 2011, MNRAS, 410, 
2662 

\bibitem[\protect\citeauthoryear{Scaife et al.}{2010}]{2010MNRAS.tmp.1771S} 
AMI Consortium: Scaife A.~M.~M., et al., 2010, MNRAS, 1771 

\bibitem[\protect\citeauthoryear{Scaife et al.}{2008}]{2008MNRAS.385..809S} 
AMI Consortium: Scaife A.~M.~M., et al., 2008, MNRAS, 385, 809 

\bibitem[\protect\citeauthoryear{Zwart et al.}{2008}]{2008MNRAS.391.1545Z} 
AMI Consortium: Zwart J.~T.~L., et al., 2008, MNRAS, 391, 1545 

\bibitem[\protect\citeauthoryear{Andre, Ward-Thompson, 
\& Barsony}{1993}]{1993ApJ...406..122A} Andre P., Ward-Thompson D., Barsony M., 1993, ApJ, 406, 122 

\bibitem[\protect\citeauthoryear{Anglada}{1996}]{1996ASPC...93....3A} 
Anglada G., 1996, ASPC, 93, 3 

\bibitem[\protect\citeauthoryear{Anglada}{1995}]{1995RMxAC...1...67A} 
Anglada G., 1995, RMxAC, 1, 67 

\bibitem[\protect\citeauthoryear{Arce 
\& Sargent}{2006}]{2006ApJ...646.1070A} Arce H.~G., Sargent A.~I., 2006, ApJ, 646, 1070 

\bibitem[\protect\citeauthoryear{Bontemps et 
al.}{1996}]{1996A&A...311..858B} Bontemps S., Andre P., Terebey S., Cabrit S., 1996, A\&A, 311, 858 

\bibitem[\protect\citeauthoryear{Boulanger et 
al.}{1996}]{1996A&A...312..256B} Boulanger F., Abergel A., Bernard J.-P., Burton W.~B., Desert F.-X., Hartmann D., Lagache G., Puget J.-L., 1996, A\&A, 312, 256 

\bibitem[\protect\citeauthoryear{Chabrier 
\& Hennebelle}{2010}]{2010ApJ...725L..79C} Chabrier G., Hennebelle P., 2010, ApJ, 725, L79 

\bibitem[\protect\citeauthoryear{Condon et al.}{1998}]{1998AJ....115.1693C} 
Condon J.~J., Cotton W.~D., Greisen E.~W., Yin Q.~F., Perley R.~A., Taylor 
G.~B., Broderick J.~J., 1998, AJ, 115, 1693 

\bibitem[\protect\citeauthoryear{Davies et al.}{2010}]{2010arXiv1012.3659D} 
Davies M.~L., et al., 2010, arXiv, arXiv:1012.3659 

\bibitem[\protect\citeauthoryear{Davis et al.}{2008}]{2008MNRAS.387..954D} 
Davis C.~J., Scholz P., Lucas P., Smith M.~D., Adamson A., 2008, MNRAS, 
387, 954 

\bibitem[\protect\citeauthoryear{de Zotti et 
al.}{2005}]{2005A&A...431..893D} de Zotti G., Ricci R., Mesa D., Silva L., Mazzotta P., Toffolatti L., Gonz{\'a}lez-Nuevo J., 2005, A\&A, 431, 893 

\bibitem[\protect\citeauthoryear{Di Francesco et 
al.}{2008}]{2008ApJS..175..277D} Di Francesco J., Johnstone D., Kirk H., 
MacKenzie T., Ledwosinska E., 2008, ApJS, 175, 277 

\bibitem[\protect\citeauthoryear{Dickinson, Davies, 
\& Davis}{2003}]{2003MNRAS.341..369D} Dickinson C., Davies R.~D., Davis R.~J., 2003, MNRAS, 341, 369 

\bibitem[\protect\citeauthoryear{Dunham et al.}{2010}]{2010ApJ...710..470D} 
Dunham M.~M., Evans N.~J., Terebey S., Dullemond C.~P., Young C.~H., 2010, 
ApJ, 710, 470 

\bibitem[\protect\citeauthoryear{Dunham et al.}{2008}]{2008ApJS..179..249D} 
Dunham M.~M., Crapsi A., Evans N.~J., II, Bourke T.~L., Huard T.~L., Myers 
P.~C., Kauffmann J., 2008, ApJS, 179, 249 

\bibitem[\protect\citeauthoryear{Emerson}{1988}]{1988felm.conf...21E} 
Emerson J.~P., 1988, felm.conf, 21 

\bibitem[\protect\citeauthoryear{Enoch et al.}{2007}]{2007ApJ...666..982E} 
Enoch M.~L., Glenn J., Evans N.~J., II, Sargent A.~I., Young K.~E., Huard 
T.~L., 2007, ApJ, 666, 982 

\bibitem[\protect\citeauthoryear{Evans et al.}{2009}]{2009ApJS..181..321E} 
Evans N.~J., et al., 2009, ApJS, 181, 321 

\bibitem[\protect\citeauthoryear{Green}{2007}]{2007BASI...35...77G} Green 
D.~A., 2007, BASI, 35, 77 

\bibitem[\protect\citeauthoryear{Harvey et al.}{2007}]{2007ApJ...663.1149H} 
Harvey P., Mer{\'{\i}}n B., Huard T.~L., Rebull L.~M., Chapman N., Evans 
N.~J., II, Myers P.~C., 2007, ApJ, 663, 1149 

\bibitem[\protect\citeauthoryear{Hatchell 
\& Dunham}{2009}]{2009A&A...502..139H} Hatchell J., Dunham M.~M., 2009, A\&A, 502, 139 

\bibitem[\protect\citeauthoryear{Hatchell et 
al.}{2007}]{2007A&A...468.1009H} Hatchell J., Fuller G.~A., Richer J.~S., Harries T.~J., Ladd E.~F., 2007, A\&A, 468, 1009 

\bibitem[\protect\citeauthoryear{Hatchell, Fuller, 
\& Richer}{2007b}]{2007A&AÉ472..187H} Hatchell J., Fuller G.~A., Richer J.~S., 2007b, A\&A, 472, 187 

\bibitem[\protect\citeauthoryear{Hatchell et 
al.}{2005}]{2005A&A...440..151H} Hatchell J., Richer J.~S., Fuller G.~A., Qualtrough C.~J., Ladd E.~F., Chandler C.~J., 2005, A\&A, 440, 151 

\bibitem[\protect\citeauthoryear{Kauffmann et 
al.}{2008}]{2008A&A...487..993K} Kauffmann J., Bertoldi F., Bourke T.~L., Evans N.~J., II, Lee C.~W., 2008, A\&A, 487, 993 

\bibitem[\protect\citeauthoryear{Kenyon 
\& Hartmann}{1995}]{1995ApJS..101..117K} Kenyon S.~J., Hartmann L., 1995, ApJS, 101, 117 

\bibitem[\protect\citeauthoryear{Kenyon et al.}{1990}]{1990AJ.....99..869K} 
Kenyon S.~J., Hartmann L.~W., Strom K.~M., Strom S.~E., 1990, AJ, 99, 869 

\bibitem[\protect\citeauthoryear{Fuller 
\& Ladd}{2002}]{2002ApJ...573..699F} Fuller G.~A., Ladd E.~F., 2002, ApJ, 573, 699 

\bibitem[\protect\citeauthoryear{Lee et al.}{2010}]{2010ApJ...709L..74L} 
Lee J.-E., et al., 2010, ApJ, 709, L74 

\bibitem[\protect\citeauthoryear{Lee et al.}{2006}]{2006ApJ...648..491L} 
Lee J.-E., et al., 2006, ApJ, 648, 491 

\bibitem[\protect\citeauthoryear{Ossenkopf 
\& Henning}{1994}]{1994A&A...291..943O} Ossenkopf V., Henning T., 1994, A\&A, 291, 943 

\bibitem[\protect\citeauthoryear{Panagia 
\& Felli}{1975}]{1975A&A....39....1P} Panagia N., Felli M., 1975, A\&A, 39, 1 

\bibitem[\protect\citeauthoryear{Paradis et 
al.}{2009}]{2009AJ....138..196P} Paradis D., et al., 2009, AJ, 138, 196 

\bibitem[\protect\citeauthoryear{Patnaik et 
al.}{1992}]{1992MNRAS.254..655P} Patnaik A.~R., Browne I.~W.~A., Wilkinson 
P.~N., Wrobel J.~M., 1992, MNRAS, 254, 655 

\bibitem[\protect\citeauthoryear{Planck Collaboration et 
al.}{2011}]{2011arXiv1101.2035P} Planck Collaboration, et al., 2011, arXiv, 
arXiv:1101.2035 

\bibitem[\protect\citeauthoryear{Reynolds}{1986}]{1986ApJ...304..713R} 
Reynolds S.~P., 1986, ApJ, 304, 713 

\bibitem[\protect\citeauthoryear{Rodr{\'{\i}}guez, Anglada, 
\& Curiel}{1999}]{1999ApJS..125..427R} Rodr{\'{\i}}guez L.~F., Anglada G., Curiel S., 1999, ApJS, 125, 427 

\bibitem[\protect\citeauthoryear{Shirley et 
al.}{2011a}]{2011AJ....141...39S} Shirley Y.~L., Mason B.~S., Mangum J.~G., 
Bolin D.~E., Devlin M.~J., Dicker S.~R., Korngut P.~M., 2011, AJ, 141, 39 

\bibitem[\protect\citeauthoryear{Shirley et 
al.}{2011b}]{2011ApJ...728..143S} Shirley Y.~L., Huard T.~L., Pontoppidan 
K.~M., Wilner D.~J., Stutz A.~M., Bieging J.~H., Evans N.~J., II, 2011, 
ApJ, 728, 143 

\bibitem[\protect\citeauthoryear{Shirley et 
al.}{2007}]{2007ApJ...667..329S} Shirley Y.~L., Claussen M.~J., Bourke 
T.~L., Young C.~H., Blake G.~A., 2007, ApJ, 667, 329 

\bibitem[\protect\citeauthoryear{Shu}{1977}]{1977ApJ...214..488S} Shu 
F.~H., 1977, ApJ, 214, 488 

\bibitem[\protect\citeauthoryear{van Kempen et 
al.}{2010}]{2010A&A...518L.121V} van Kempen T.~A., et al., 2010, A\&A, 518, L121 

\bibitem[\protect\citeauthoryear{Walawender, Bally, 
\& Reipurth}{2005}]{2005AJ....129.2308W} Walawender J., Bally J., Reipurth B., 2005, AJ, 129, 2308 

\bibitem[\protect\citeauthoryear{Waldram et 
al.}{2010}]{2010MNRAS.404.1005W} Waldram E.~M., Pooley G.~G., Davies M.~L., 
Grainge K.~J.~B., Scott P.~F., 2010, MNRAS, 404, 1005 

\bibitem[\protect\citeauthoryear{Wu et al.}{2007}]{2007AJ....133.1560W} Wu 
J., Dunham M.~M., Evans N.~J., II, Bourke T.~L., Young C.~H., 2007, AJ, 
133, 1560 

\bibitem[\protect\citeauthoryear{Young 
\& Evans}{2005}]{2005ApJ...627..293Y} Young C.~H., Evans N.~J., II, 2005, ApJ, 627, 293 

\bibitem[\protect\citeauthoryear{Young et al.}{2004}]{2004ApJS..154..396Y} 
Young C.~H., et al., 2004, ApJS, 154, 396 




\end{thebibliography}
\end{document}